\documentclass[11pt, draftclsnofoot, peerreview, onecolumn]{IEEEtran}

\IEEEoverridecommandlockouts

\usepackage{lineno}
\usepackage{amssymb}
\usepackage{amsmath}
\usepackage{amsthm}
\usepackage{epsfig}
\usepackage{graphicx}
\usepackage{graphics}
\usepackage{float}
\usepackage{subfigure}
\usepackage{multirow}
\usepackage{color}
\usepackage{fullpage}
\usepackage[normalem]{ulem}
\usepackage{makeidx}
\usepackage{xspace}
\usepackage{wrapfig}
\usepackage{lipsum}
\usepackage{mathtools}
\usepackage{cuted}
\usepackage{cite}
\usepackage{amsfonts}
\usepackage{algorithmic}
\usepackage{textcomp}
\usepackage{amsmath}     % For math environments and symbols
\usepackage{amssymb}     % For additional math symbols
\usepackage[T1]{fontenc}      % safe text encoding
\usepackage{newtxtext,newtxmath} % Times-like text+math, arXiv-friendly

\usepackage[ruled,vlined]{algorithm2e}

\makeindex
\theoremstyle{plain}
\newtheorem{theorem}{{ Theorem}}

\theoremstyle{definition}
\newtheorem{Definition}{Definition}
\newtheorem{corollary}{Corollary}

\newtheorem{exmp}{Example}
\newtheorem{remark}{Remark}
\newtheorem{lemma}{Lemma}

\makeatletter
\renewcommand\@endtheorem{\vvv@endmarker\endtrivlist\@endpefalse}
\newcommand\vvv@endmarker{%
  {\unskip\nobreak\hfil\penalty50
  \hskip2em\vadjust{}\nobreak\hfil\openbox
  \parfillskip=0pt \finalhyphendemerits=0 \par
  \penalty 10000 \parskip=0pt\noindent}\ignorespaces}
\makeatother

\definecolor{darkred}{rgb}{1, 0.1, 0.3}
\definecolor{darkblue}{rgb}{0.1, 0.1, 1}
\definecolor{darkgreen}{rgb}{0,0.6,0.5}
\def\BibTeX{{\rm B\kern-.05em{\sc i\kern-.025em b}\kern-.08em
    T\kern-.1667em\lower.7ex\hbox{E}\kern-.125emX}}

\DeclareMathOperator{\hyperminrank}{hyper-minrank}

\DeclareMathOperator{\minrk}{minrk}
\DeclareMathOperator{\rank}{rank}
\newcommand{\F}{\mathbb{F}}

\def \A {\mathbf{A}}
\def \a {\mathbf{a}}
\def \G {\mathcal{G}}

\def \B {\mathbf{B}}

\def \X {\mathcal{X}}
\def \V {\mathcal{V}}

\def \I {\mathcal{I}}
\def \J {\mathcal{J}}
\def \F {\mathbb{F}}

\def \M {\mathcal{M}}
\def \e {\mathbf{e}}

\def \c {\mathbf{c}}
\def \u {\mathbf{u}}
\def \x {\mathbf{x}}
\def \E {\mathcal{E}}
\def \Supp {\text{supp}}

\def \rank {\text{rank}}
\allowdisplaybreaks

\begin{document}
\cleardoublepage
\setcounter{page}{1}
\title{{ Hyper-Minrank: A Unified Hypergraph Characterization of Multi-Sender Index Coding}}

\author{Ali~Khalesi,~\textit{Member,~IEEE},~and~Petros~Elia,~\textit{Member,~IEEE}%
\thanks{%
This work was supported by the European Research Council (ERC) through the EU Horizon 2020 Research and Innovation Program under the ERC Proof-of-Concept project LIGHT (Grant 101101031), the Huawei France-funded Chair towards Future Wireless Networks, and by the French government under the France 2030 ANR program “PEPR Networks of the Future” (ref. ANR-22-PEFT-0010). 
A.~Khalesi is with the Signals \& Artificial Intelligence (SIA) Department, IPSA – Institut Polytechnique des Sciences Avancées, Ivry-sur-Seine, France (e-mail: \texttt{ali.khalesi@ipsa.fr}). 
P.~Elia is with the Communication Systems Department, EURECOM, 450 Route des Chappes, 06410 Sophia Antipolis, France (e-mail: \texttt{elia@eurecom.fr}).%
}}
\maketitle

\begin{abstract}
This work introduces a hypergraphic formulation that \emph{generalizes} the classical paradigm of Bar-Yossef et al. to the \emph{multi-sender} index coding (MSIC) setting. Key to this hypergraphic model is a side-information \emph{$4$-regular hypergraph} $\mathcal{G}$, a new adjacency representation $\mathbf{A}_{\mathcal{G}}=[\mathbf{A}_1\,\cdots\,\mathbf{A}_N]$, and a simple fitting criterion for sub-hypergraph validity, all in the presence of specially designed hyperedges that capture side information and cross-sender signal cancellation. Interestingly, this formulation establishes a tight achievability--converse equivalence for the entire $N$-sender $K$-receiver problem: every valid fitting induces a valid linear multi-sender index code, every linear code induces a valid fitting, and the optimal scalar linear broadcast length equals the \emph{hyper-minrank} $\ell^\star_{\mathrm{lin}}(\mathcal{G})=\hyperminrank(\mathcal{G})\triangleq \min_{\mathbf{A}\text{ fits }\mathcal{G}} \sum_{n=1}^N \rank(\mathbf{A}_n)$. As a consequence, beyond this exact characterization, our approach offers  \emph{hypergraph} analogues of Haemers-type bounds on the broadcast length: a clique-cover upper bound, and a lower bound via the clique number of a carefully defined complement hypergraph. 
Algorithmically, we provide an exact procedure to compute the optimal $\hyperminrank(\mathcal{G})$, while showing that for certain regimes, its complexity is asymptotically better than approximate LT–CMAR solutions.
The framework captures well-known settings like embedded index coding, while applying directly to a variety of settings such as multi-sender cache-aided communications, coded computation, distributed storage, and edge/satellite systems, which can now accept  $\hyperminrank$ as a unified design target.
\end{abstract}
\section{Introduction}
The \emph{index--coding problem} studies how a broadcaster can exploit side information at the receivers, to accelerate the delivery of desired files via multicast transmissions.
The idea originated with the “informed--source coding--on--demand” (ISCOD) scheme of Birk and Kol~\cite{BirkKol1998,birk2006coding}, motivated by its applications to satellite communications, and was later recast in graph-theoretic form by Bar-Yossef \emph{et al.}, who showed that the \emph{minrank} of the side-information graph exactly characterizes the shortest scalar linear code length~\cite{Bar-Yossef1}.

Focusing on this single-sender setting, a large body of research has examined \emph{graph-theoretic approaches} in an effort to identify the optimal broadcast rate.  
Key to these approaches is the idea that the side-information structure can be modeled by a directed graph whose vertices represent receivers and whose edges represent known packets.  
Based on this, many of the early foundational results linked the optimal scalar linear code length to the \emph{minrank} of this side-information graph~\cite{Bar-Yossef1,Haemers2006,Golovnev1,Alon2018TheMO,Haviv2018minrank,10.1145/3322817}, prompting extensive study of minrank behavior for special graph classes as well as for random graphs~\cite{Haviv5,Coja2005lovasz}.  
This graph-theoretic take on the problem provided a variety of clever graph-coloring-based upper bounds, such as clique cover and cycle cover bounds~\cite{Shanmugam2013local,Shanmugam2014graph,Tahmasbi2014critical}. These same approaches were also seen as more computationally tractable surrogates for minrank, and were often refined using \emph{local chromatic number} arguments~\cite{Shanmugam2013local} that exploit neighborhood structures to reduce code length.  
Along this line of research, the related \emph{interlinked-cycle cover} approach of Thapa \emph{et al.}~\cite{Thapa2017interlinked} generalized both cycle and clique covers by exploiting overlapping cycles in the graph, yielding strictly tighter bounds in certain topologies, while in terms of converses, confusion graph techniques~\cite{Arbabjolfaei1} and their fractional chromatic numbers were used to prove information-theoretic lower bounds on broadcast rate, and were later adapted to other settings such as two-sender and embedded index coding~\cite{Thapa,Porter1}.  

A distinct but related line of work considers the \emph{embedded index coding} problem, which can be seen as a distributed variant of index coding where typically nodes act both as senders and receivers of information~\cite{Porter1,9080099,Sundar3}.  
This approach --- which carries many applications such as in distributed storage systems, coded computation, and vehicular networks --- was first formalized by Porter and Wootters in~\cite{Porter1} that provided general bounds, connecting embedded instances to constrained versions of the classical problem.
Subsequently, Haviv~\cite{9080099} {\color{black}investigated} the concept of \emph{task-based solutions} in which the coding design is optimized jointly over the embedded index coding constraints and the overarching system objectives. Additionally, Mahesh \emph{et al.}~\cite{Sundar3} studied the \emph{minrank} of embedded problems, providing characterizations for various classes of structured side information.  In the end, many of these works reveal how topology introduces additional structural idiosyncratic properties—e.g., locality, ordering, or partial message availability—that fundamentally alter both achievability and converse arguments compared to the original unconstrained index coding problem. 

Continuing along the same path, the \emph{multi-sender index coding} problem—also known as \emph{distributed index coding}— was introduced in~\cite{ong2016single} as an extension of the classical single-sender problem, where now multiple senders (or senders), each store an arbitrary subset of the message library, and where each sender connects to all receivers via a noiseless broadcast links of limited capacity.  This is a key setting as it captures crucial scenarios --- in communications, computing and learning --- characterized by having no single point of aggregation for all messages. This defining characteristic introduces a new class of challenges, and the task of jointly designing transmissions across senders to minimize the total broadcast load becomes substantially more involved compared to the original index coding case.  
As stated, the work in \cite{ong2016single} first introduced the multi-sender formulation, extending the single-uniprior model and deriving structural properties and exact capacities for certain classes of problems.  
Subsequently, Ong, Lim, and Ho~\cite{ong2013multi} provided achievable schemes as well as shed light on the interplay between message distribution and side information.  Furthermore, Li, Ong, and Johnson~\cite{li2017improved} refined existing bounds through interference alignment arguments, while~\cite{li2018cooperative,li2018multi} introduced cooperative compression and rank-minimization techniques to better exploit overlapping message sets. In addition, Thapa \emph{et al.}~\cite{Thapa} analyzed the two-sender unicast index coding problem (TSUIC), decomposing it into independent subproblems and introducing the notion of two-sender graph coloring to characterize optimal rates, while Arunachala \emph{et al.}~\cite{Rajan7,8836660} identified TSUIC classes with \emph{fully-participated interactions}, i.e., cases where every sender participates in the exchange of all messages involved in each interaction, such that reducing side information or pruning links does not alter the optimal broadcast rate. { Additionally, Ghaffari~\emph{et al.}~\cite{Ghaffari44} studied multi-sender index coding over linear networks, developing bounds under the effect of interference\footnote{{ This work considered a modified framework by introducing a linear network mixing model. Unlike in the standard setting, where each sender transmits independently to all receivers over separate noiseless links, in this model, the transmitted signals are linearly combined by the network before reaching the receivers. As a result, each receiver obtains a mixture of all senders’ transmissions rather than distinct streams, effectively capturing both wireless interference and wired network-coding scenarios.
}}, }while Kim and No~\cite{Kim1} proposed a fitting-matrix framework for linear multi-sender index coding in cellular networks.  Furthermore,  Sadeghi, Arbabjolfaei, and Kim~\cite{sadeghi2016distributed} characterized capacity regions for small distributed index coding instances via composite coding schemes and cut-set bounds, while Liu \emph{et al.}~\cite{Sadeghi2} derived general converse bounds and achievable schemes for arbitrary message-server assignments, and Sadeghi \emph{et al.}~\cite{Sadeghi3} obtained exact capacities for broader classes, including those with symmetry-exploiting side information.  

To better explain our approach here, it is now worth offering a small summary of the main ideas of the seminal work in~\cite{Bar-Yossef1}, as well as some insight on how we will extend these ideas to the multi-sender case.

\paragraph{Linear single-sender model via side-information graphs}
In~\cite{Bar-Yossef1}, the work considers a directed side-information digraph $G=({[K]},\mathcal{E})$ with a vertex set $[K] \triangleq \{1,2,\hdots,K\}$ and an edge set $\mathcal{E}$ that reflects how each receiver $k\in[K]$ wishes to receive $x_k$ while knowing $\{x_{k'}: (k\to k')\in \mathcal{E}\}$. In this context, a binary matrix $\mathbf{M}\in\F_2^{K\times K}$ is said to \emph{fit} $G$ if
\begin{equation}
\mathbf{M}(k,k)=1\quad \forall k\in[K], 
\qquad
\mathbf{M}({k,k'})=0\quad \text{whenever } (k\to k')\notin \mathcal{E}
\label{eq:fit}
\end{equation}
with the rest of the entries remaining free, which in turn brings about the so-called \emph{fitting family} which takes the form
\begin{equation}
\mathcal{M}(G)\;\triangleq\;\bigl\{\,M\in\F_2^{K\times K}: M \text{ satisfies \eqref{eq:fit}}\,\bigr\}
\end{equation}
The resulting main result in~\cite{Bar-Yossef1} is that the \emph{minrank} of $G$ over $\F_2$, which takes the form 
\begin{equation}
\minrk_2(G)\;\triangleq\;\min_{\mathbf{M}\in\mathcal{M}(G)} \rank_{\F_2}(M)
\label{eq:minrk-def}
\end{equation}
in fact characterizes the exact optimal scalar linear index codelength $\ell^\star_{\mathrm{lin}}(G)$, as follows  
\begin{equation}
\ell^\star_{\mathrm{lin}}(G)\;=\;\minrk_2(G)
\label{eq:byjk-thm1}
\end{equation}
Consequently, any $\mathbf{M}^\star\in\mathcal{M}(G)$ attaining $\minrk_2(G)$ yields an optimal code whose encoder consists of any $\minrk_2(G)$ linearly independent rows of $\mathbf{M}^\star$, as well as yields the corresponding decoders that recover $x_i$ by combining the received transmissions with the local side information \cite{Bar-Yossef1}.

\paragraph{Classical graph-theoretic bounds for index coding}
For an \emph{undirected} graph $G$, minrank enjoys Haemers-type graph theoretic bounds that lie between the Shannon capacity $\Theta(G)$ and the chromatic number $\chi(\bar{G})$ of the complement graph $\bar{G}$:
\begin{equation}
\omega(\bar{G})\;\le\;\Theta(G)\;\le\;\minrk_2(G)\;\le\;\chi(\bar{G}) = \theta(G)
\label{eq:haemers-sandwich}
\end{equation}
which in turn are bounded by the clique cover number $\theta(G)$ and the clique number $\omega(\bar{G})$. These inequalities serve as bounds on $\ell^\star_{\mathrm{lin}}$ (cf.~\eqref{eq:byjk-thm1}, see~\cite[Prop.~1]{Bar-Yossef1}).

\begin{figure}
    \centering
\includegraphics[width=0.8\linewidth]{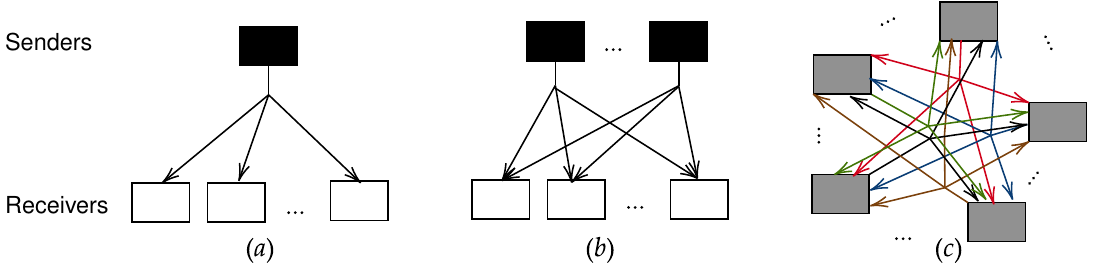}
\vspace{-10pt}
    \caption{{Communication models: (a) centralized index coding with a single sender and multiple receivers; 
(b) general multi-sender index coding with multiple senders and  receivers; 
(c) embedded index coding, a special case of (b) where each sender is also a receiver, yielding multiple joint sender–receiver nodes.}
}
    \label{All-Problems}
\end{figure}
\noindent\textbf{Our results (From single sender to multiple senders)}
We now provide a brief overview of our approach, which builds on the minrank framework in the single-sender setting. We first introduce a \emph{hypergraphic} representation of the linear multi-sender index-coding problem, which natively captures the potential for cross-sender cancellations as a function of message placement across multiple senders, while also capturing the demands and the placement of the data blocks at each sender. As with most related works, we focus on the single-unicast case, where each receiver demands a single data block (and each data block is demanded by a single receiver), as well as focus on the binary field case. The defined directed side-information hypergraph $\mathcal{G}(\mathcal{V}, \mathcal{E})$, will be such that
\[
\mathcal{V} \subseteq \{\, k, \mathcal{S}\mid k \in \{1,2,\hdots, K\},~ \mathcal{S} \subseteq \{1,2,\hdots,N\},~ |\mathcal{S}| \leq 2\,\}
\]
where $k$ represents the receiver–data-block index and $\mathcal{S}$ represents the (one or two) senders involved. Even though there are multiple senders, our focus on having $ |\mathcal{S}| \leq 2$ in our representation will prove non-restrictive. An important point is that, the hyperedge set $\mathcal{E}$ will consist of three disjoint classes:
    \begin{enumerate}
  \item \textbf{Class 1: Demand hyperedges} $\mathcal{E}_d$. These hyperedges take the form $(\{k\},\{k,\{n\}\})$ and represent the case where the $k$-th data block, requested by receiver $k$, is stored at sender $n$. While in the single-sender case, such edges were unnecessary since the unique sender had all blocks, in the multi-sender case, these edges are key to certifying code validity. In the $4$-uniform representation, such a demand hyperedge will be represented by a 4-tuple of the form $(k,k,n,n)$.
    \item \textbf{Class 2: Cached-data hyperedges} $\mathcal{E}_s$. These hyperedges take the form $(\{k\},\{k',\{n\}\})$ and represent the case where the $k'$-th data block is stored at receiver $k$ and sender $n$. This can be seen as a direct extension of the directed edge of a single-sender side-information graph. Such an edge will have a $4$-uniform representation of $(k,k',n,n)$.
    \item \textbf{Class 3: Coupled hyperedges} $\mathcal{E}_c$. These hyperedges take the form $(\{k\},\{k',\{n,n'\}\})$ and represent the case where receiver $k$ has not cached the $k'$-th data block, which is stored though at senders $n$ and $n'$. The emergence of such edges is specific to the multi-sender setting, as they capture cancellation of undesired contributions via linear combinations of transmissions from \emph{multiple} senders. Such edge will have a $4$-uniform representation of $(k,k',n,n')$.
\end{enumerate}

For example, a class-1 edge $(1,1,2,2)$ indicates that the first data block requested by receiver~1 is stored at sender~2. 
A class-2 edge $(1,3,4,4)$ specifies that the third data block is stored by receiver~1 and is also available at sender~4. 
Finally, a class-3 edge $(1,5,4,6)$ denotes that file~5 is accessible from senders~4 and~6 but is not stored by receiver~1.

As there is no universal adjacency notion for hypergraphs, we will introduce a \emph{composite adjacency matrix} tailored to this problem, enabling us to capture all linear index codes. This matrix 
\[
\mathbf{A}_{\mathcal{G}}=[\mathbf{A}_1~\mathbf{A}_2~\cdots~\mathbf{A}_N] \in \mathbb{F}^{K \times KN}
\] is written as the concatenation of $N$ blocks 
$\mathbf{A}_{n} \in \mathbb{F}^{K \times K}$, one per sender.  
Each block $\mathbf{A}_{n}$ will dictate (encode) which message symbols stored at 
sender~$n$ may participate in the linear combinations that aid each 
receiver~$k$, thereby determining the allowable encoding actions of 
sender~$n$.  
Altogether, the matrix $\mathbf{A}_{\mathcal{G}}$ summarizes the 
full set of sender–receiver linear dependencies induced by the 
hypergraph~$\mathcal{G}$.  
Its formal construction appears in Definition~\ref{Adjacency matrix-def}.
In the single-sender case, this reduces to the classical adjacency of the directed side-information graph. A key point worth noting is that unlike in the single-sender case, now matrices whose non-diagonal non-zero entries are supported within $\mathbf{A}_{\mathcal{G}}$ do not necessarily represent valid multi-sender codes. We therefore refine \emph{fitting}: we will consider conditions for a matrix $\mathbf{A}'$ to \emph{fit} $\mathcal{G}$, and connect this to the notion of \emph{valid} sub-hypergraphs of $\mathcal{G}$. Unlike in the single-sender case, not every sub-hypergraph is valid; to be valid, it must satisfy a simple condition, which is that each receiver vertex $k$ participates in an \emph{odd} number of demand hyperedges. As we will see, this condition guarantees correctness and yields that
\[
\sum_{n \in [N]}\rank(\mathbf{A}'_{n})
\]
is a lower bound on the index-code length. To obtain the optimal length, we introduce a hypergraphic generalization of the \emph{minrank} function:
\begin{align}
\hyperminrank(\mathcal{G})~\triangleq~\min \Big\{\,\sum^{N}_{n=1} \rank(\mathbf{A}_n)\,\Big|\,\mathbf{A}\ \text{fits}\ \mathcal{G}\Big\}
\end{align}

\noindent Then, in Theorem~1, we show that for any side-information hypergraph, there exists a linear multi-sender index code whose length equals $\hyperminrank(\mathcal{G})$, and that no linear multi-sender index code can be shorter.

{
To prove the above, we will follow the general approach of Theorem~1 in~\cite{Bar-Yossef1}. 
Specifically, we show that the spanning submatrices of $\mathbf{A}'_n$, $n \in [N]$ --- where $\mathbf{A}'$ 
is the composite adjacency matrix of a valid sub-hypergraph of $\mathcal{G}$ --- correspond to a valid index code, 
in the sense that the columns of each $\mathbf{A}'_n$ define the \emph{encoding vectors} used by sender~$n$. 
Each sender can thus form its transmitted linear combination by weighting its locally stored messages according to these column vectors. 
Conversely, we show that any arbitrary multi-sender linear index code induces the composite adjacency matrix of a valid sub-hypergraph of $\mathcal{G}$, 
with the corresponding $\sum_{n \in [N]} \rank(\mathbf{A}'_n)$ lower-bounding the total code length.}

A key challenge lies in properly accounting for \emph{demand} and \emph{coupled} hyperedges. 
Demand edges ensure that each desired block is delivered with odd parity, thereby ascertaining the validity of the sub-hypergraph. 
Coupled edges, on the other hand, capture a phenomenon unique to the multi-sender setting: a receiver can cancel undesired symbols by combining transmissions across multiple senders --- something that a metric such as the ``sum of $N$ single-sender minranks'' cannot capture\footnote{Such a metric would capture optimality, only under the assumption that sender message sets are disjoint~\cite{Thapa}.}. 
For example, if receiver~1 has $x_3$, and if sender~1 transmits $x_1 + x_2 + x_3$ while sender~2 transmits $x_2$, then $x_1$ can be decoded by summing the two broadcast signals and removing $x_3$ at the receiver. 
As we will see, demand, cached, and coupled edges will together capture \emph{all} necessary linear encoding/decoding strategies.

\paragraph*{Bounds}
With $\mathcal{G}$ and $\hyperminrank(\mathcal{G})$ in place, we will extend classical graph-theoretic bounds (cf.~\eqref{eq:haemers-sandwich}) to our case. In this context, in Theorem~2 we show that identifying which hypergraphs correspond to a single transmission in the multi-sender code leads to a notion of (valid) hypergraphic cliques. Then, we cover $\mathcal{G}$ by such valid cliques $\mathcal{C}_1,\dots,\mathcal{C}_m$, i.e., we choose cliques such that
\[
\{\, k,k' \mid (k,k',n,n') \in \cup_{t \in [m]} \mathcal{C}_t \,\} = [K]
\]
and such that the collection of hypergraphic cliques $\{\mathcal{C}_j\}_{j=1}^m$ jointly forms a valid sub-hypergraph (their union $\bigcup_{j=1}^{m}\mathcal{C}_j$ satisfies the validity conditions of Definition~\ref{Valid sub-hypergraph}). This construction implies
\[
\hyperminrank(\mathcal{G}) \leq m.
\]
Then, for a complementary lower bound, in Theorem~3 we show that by defining a suitable complement hypergraph $\bar{\mathcal{G}}$ (complement of $\mathcal{G}$) and identifying its largest admissible clique $\mathcal{C}$, we obtain
\[
|\mathcal{V}(\mathcal{C})| \leq \hyperminrank(\mathcal{G})
\]
where $\mathcal{V}(\mathcal{C})$ denotes the set of vertices participating in the clique $\mathcal{C}$ of the complement $\bar{\mathcal{G}}$.
\iffalse
With $\mathcal{G}$ and $\hyperminrank(\mathcal{G})$ in place, we will extend classical graph-theoretic bounds (cf.~\eqref{eq:haemers-sandwich}) to our case. In this context, in Theorem~2 we show that identifying which hypergraphs correspond to a single transmission in the multi-sender code leads to a notion of (valid) hypergraphic cliques. Then, covering $\mathcal{G}$ by such valid cliques $\mathcal{C}_1,\dots,\mathcal{C}_m$  ---  i.e., making sure that
\[
\{\, k,k' \mid (k,k',n,n') \in \cup_{t \in [m]} \mathcal{C}_t \,\} = [K]
\]

where the collection of hypergraphic cliques $\{\mathcal{C}_j\}_{j=1}^m$ jointly forms a valid sub-hypergraph, 
i.e., their union $\bigcup_{j=1}^{m}\mathcal{C}_j$ satisfies the validity conditions of Definition~\ref{Valid sub-hypergraph}. The above implies
\[
\hyperminrank(\mathcal{G}) \leq m.
\] and eventually yields
\[
\hyperminrank(\mathcal{G}) \leq m.
\]
Then, for a complementary lower bound, in Theorem~3 we show that by defining a suitable complement hypergraph $\bar{\mathcal{G}}$ (complement of $\mathcal{G}$) and identifying its largest admissible clique $\mathcal{C}$, we obtain
\[
|\mathcal{V}(\mathcal{C})| \leq \hyperminrank(\mathcal{G})
\]
where $\mathcal{V}(\mathcal{C})$ denotes the set of vertices participating in the clique $\mathcal{C}$ of the complement $\bar{\mathcal{G}}$.
\fi
\paragraph*{Algorithm and complexity}
We also present an exact algorithm for computing $\hyperminrank(\mathcal{G})$ and establish its correctness. 
The algorithm’s search space is shown to be no larger than that of the optimal method in~\cite{Kim1}, 
while also outperforming the heuristic LT--CMAR~\cite{li2018multi} in computational efficiency and precision. 
For embedded index coding~\cite{Porter1}, we further quantify the computational overhead, of our approach, associated with achieving exact optimality.

\medskip
\noindent\textbf{Summary of main contributions:}
\begin{itemize}
\item \textbf{Optimal linear scheme via hyper-minrank.}
We design an optimal linear multi-sender index-coding scheme by introducing the \emph{hyper-minrank} functional $\hyperminrank(\mathcal{G})$ on a constructed side-information hypergraph $\mathcal{G}$, and we prove, via achievability and converse, that the optimal linear broadcast length equals $\hyperminrank(\mathcal{G})$.

\item \textbf{Extension of the minrank framework.}
Building on the work in~\cite{Bar-Yossef1}, we lift the single-sender fitting/minrank characterization to the multi-sender setting, thereby generalizing the classical theory from graphs to hypergraphs.

\item \textbf{Novel formulation: $4$-regular hypergraphic representation.}
We model the multi-sender index coding problem by a directed side-information \emph{hypergraph} whose hyperedges are represented by 4-tuples $(k,k',n,n')$, capturing (i) demands, (ii) cached side information, and (iii) \emph{coupled} interactions across senders.
\begin{itemize}
\item \textbf{Coupled hyperedges for inter-sender cancellation.}
We introduce \emph{coupled hyperedge sets} that explicitly encode the receiver-side ability to cancel undesired symbols by forming linear combinations \emph{across multiple senders}. %, a phenomenon absent in the single-sender model.

\item \textbf{Adjacency and fitting for code construction.}
We define a block composite adjacency matrix $\mathbf{A}_{\mathcal{G}}=[\mathbf{A}_1~\cdots~\mathbf{A}_N]$ and a notion of \emph{valid} sub-hypergraph fitting. 
For any valid fitting $\mathbf{A}'$, the encoder at each sender is obtained from spanning submatrices of the corresponding block $\mathbf{A}'_n$. Conversely, any linear multi-sender index code induces a valid fitting, yielding $\sum_{n=1}^N \rank(\mathbf{A}'_n)$ as a fundamental lower bound.
\end{itemize}
\item \textbf{Exactness despite apparent restrictiveness.}
Although our hypergraphic formalism may appear restrictive, our converse shows it is tight: the induced $\hyperminrank(\mathcal{G})$ exactly characterizes the optimal scalar linear performance, thus guaranteeing no loss relative to the best linear code.

\item \textbf{Graph theoretic bounds in the multi-sender setting.}
We extend Haemers-style graph-theoretic bounds to hypergraphs, which yields a clique-cover upper bound and a complementary clique-number lower bound for $\hyperminrank(\mathcal{G})$ (Theorems~2 and~3), paralleling the classical graph-theoretic bounds.

\item \textbf{Algorithmic contributions and complexity guarantees.}
We present an exact algorithm for computing the desired \(\hyperminrank(\mathcal{G})\) and for establishing its correctness. We explicitly characterize the search-space size, and we show that in general multi-sender settings, our algorithm’s search space is never larger than that of the method in~\cite{Kim1}, while we further benchmark against the generally suboptimal heuristic LT--CMAR~\cite{li2018multi}, identifying a practical regime where our optimal method is asymptotically more efficient. 
\item \textbf{Embedded index coding trade-offs.}
For the embedded setting~\cite{Porter1}, we quantify the additional computational cost required to attain the exact optimum, clarifying the trade-off between precision and savings in approximate schemes.
\end{itemize}

\noindent\textbf{Paper organization.}
Section~\ref{sec1} formalizes the multi-sender single-unicast model and introduces the directed side-information hypergraph (Definition~\ref{side-info-graph}), the hypergraphic composite adjacency matrix (Definition~\ref{Adjacency matrix-def}), and the notion of valid sub-hypergraphs (Definition~\ref{Valid sub-hypergraph}). The same section then defines $\hyperminrank(\mathcal{G})$ and proves the achievability–converse equivalence (Theorem~\ref{main-Theorem}), concluding with a simple example.  
Section~\ref{sec2} defines hypergraphic cliques (Definition~\ref{Hypergraphic-clique}) and valid cliques (Definition~\ref{Valid-Clique}), and presents a clique-cover upper bound (Theorem~\ref{Theorem-upper}) and a subsequent lower bound (Theorem~\ref{Theorem-lower}). 
Section~\ref{sec3} presents an exact algorithm to compute $\hyperminrank(\mathcal{G})$. %, and analyzes this algorithm. %the optimal exhaustive method of~\cite{Kim1} and the LT--CMAR heuristic of~\cite{li2018multi}, including a specialization to embedded index coding~\cite{Porter1}.  
Section~\ref{sec4} concludes by discussing potential applications and future research, while Appendix~\ref{Example} provides a detailed illustrative example that takes us through all key components of the framework, including hypergraph construction, valid sub-hypergraphs, adjacency matrices, and the subsequent clique-based bounds. %The example demonstrates concretely how the proposed definitions and theorems operate on a nontrivial instance of the problem.

\paragraph*{Notations}
We use $[n]\triangleq\{1,2,\ldots,n\}$. Unless stated otherwise, all arithmetic is over the binary field $\F_2$, and we use $\rank(\cdot)$ to denote rank over $\F_2$. Bold lowercase letters (e.g., $\x$) denote column vectors, bold uppercase letters (e.g., $\A$) denote matrices, while calligraphic letters (e.g., $\mathcal{G}$) denote sets or hypergraphs.
For matrices $\A,\B$, the horizontal concatenation is $[\A,\B]$. For any $\X\in\F^{m\times n}$, $\X(i,j)$ is the $(i,j)$ entry, $\X(i,:)$ the $i$th row, and $\X(:,j)$ the $j$th column. For index sets $\I\subseteq[m]$, $\J\subseteq[n]$, $\X(\I,\J)$ is the submatrix with rows in $\I$ and columns in $\J$. We write $\Supp(\X)\subseteq[m]\times[n]$ for the support of $\X$ (indices of nonzero entries). The standard basis vector is $\e_k$, the all-ones (resp. all-zeros) vector is $\mathbf{1}_m$ (resp. $\mathbf{0}_m$), the $m\times m$ identity matrix is $\mathbf{I}_m$, and $\mathbf{1}_{m\times n}$ is the $m\times n$ all-ones matrix. The indicator function is denoted as $\mathbb{1}(\cdot)\in\F$, and for a set $\mathcal{S}\subseteq[K]$, The indicator $\mathbf{1}(\mathcal{S})\in\F^{K}$ denotes its indicator column vector\footnote{For any subset $\mathcal{S}\subseteq[K]=\{1,2,\dots,K\}$, The indicator $\mathbf{1}(\mathcal{S})\in\mathbb{F}^{K}$ denotes its indicator column vector, whose $i$-th entry equals~1 if $i\in\mathcal{S}$ and~0 otherwise, i.e., $[\mathbf{1}(\mathcal{S})]_i=1$ if $i\in\mathcal{S}$ and $0$ otherwise. For example, if $K=5$ and $\mathcal{S}=\{2,4\}$, then $\mathbf{1}(\mathcal{S})=[0,1,0,1,0]^{\top}$.}.

\paragraph*{Hypergraph notations } Our problem will be linked to a directed side-information hypergraph which will be denoted as $\G=(\V,\E)$ where \[\mathcal{V} =\{k,\mathcal{S}| k \in [K], \mathcal{S} \subset [N], |\mathcal{S}| \leq 2\}\] and where its directed hyperedges take the form $(\{k\}, \{k' ,\mathcal{S}\}), k,k' \in [K], \mathcal{S} \subseteq [N], |\mathcal{S}| \leq 2$ which though, for simplicity, is henceforth denoted using 4-tuples $$(k,k',n,n')\in[K]\times[K]\times[N]\times[N]$$ where $n,n' \in \mathcal{S}$. We will later define $\E_d,\E_s,\E_c$ to be the set of demand, cached data, and coupled hyperedges, respectively.  Based on this, we will use the notation $\E=\E_d\cup\E_s\cup\E_c$ to represent the (disjoint) union of demand, cached data, and coupled hyperedges.  We will also use the notation $\mathcal{E}_{k} \triangleq \{(k,k',n,n') |(k,k',n,n') \in \mathcal{G} \}, k \in [K]$ to represent hyperedges whose starting vertex is $k$, while we will reserve notation $\mathcal{E}_{d,k}$, $\mathcal{E}_{s,k}$, $\mathcal{E}_{c,k}$ to respectively represent demand, cached data, and coupled hyperedges whose initial vertex is $k$.  Finally, for a set $\mathcal{C}$ of 4-tuples we use the projections
$\mathcal{C}_R\triangleq\{(k,k'): (k,k',n,n')\in\mathcal{C}\}$,
$\mathcal{C}_S\triangleq\{\{n,n'\}: (k,k',n,n')\in\mathcal{C}\}$, and
$\mathcal{C}_{S_n}\triangleq\{(k,k'): (k,k',n,n')\in\mathcal{C}\}$.

\section{Hypergraphic Problem Formulation and Optimal Multi-Sender Index Code}\label{sec1}
\begin{figure}
    \centering
\includegraphics[width=0.5\linewidth]{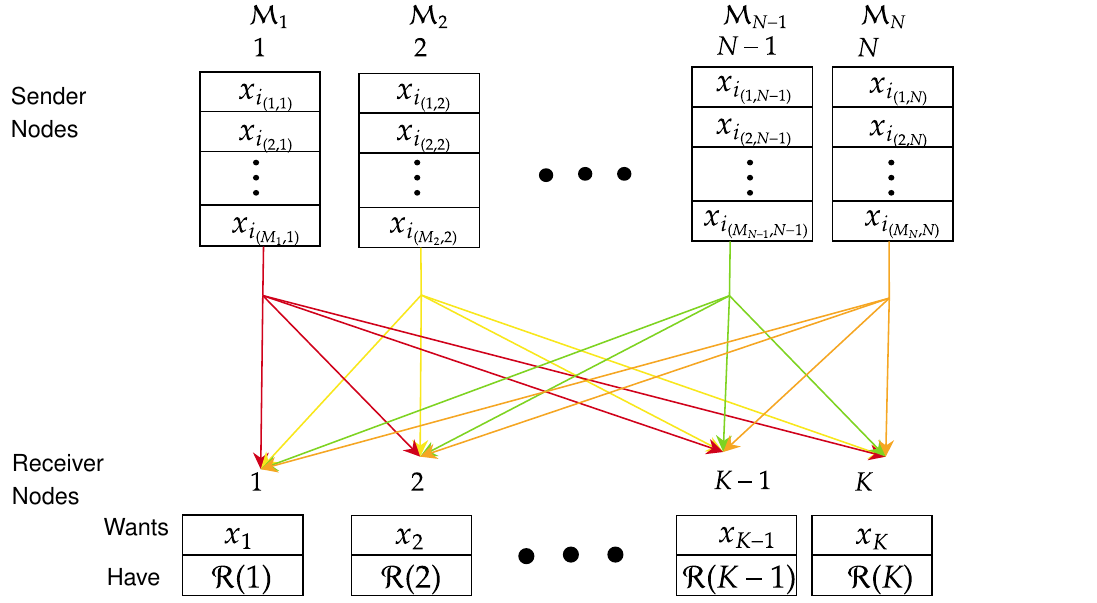}
\vspace{-10pt}
    \caption{{Multi-sender index-coding model with $N$ senders and $K$ receivers. 
Each sender $n\in[N]$ stores the subset of binary messages $\{x_i:i\in\mathcal{M}_n\}\subseteq\{x_1,\ldots,x_K\}$ and transmits over error-free, mutually non-interfering broadcast links to all receivers. 
Receiver $k\in[K]$ demands $x_k$ and knows side information $\mathcal{R}(k)\subseteq[K]\setminus\{k\}$. %Here $[N]\!\triangleq\!\{1,\ldots,N\}$ and $[K]\!\triangleq\!\{1,\ldots,K\}$.
}}
    \label{fig:enter-label}
\end{figure}

We consider the setting where there are $N$ senders and $K$ unit-sized block messages\footnote{This assumption captures the setting of block-level erasures where any node either possesses an entire error-free block or none of it. This nicely represents the practical situations where block error detection or correction is used. Given this assumption and the fact that we work on a binary field $\F$, it suffices to represent each file with a bit.} represented by $\mathbf{x} =[x_1,x_2,\hdots,x_{K}] \in \F^{K}$. Sender $n$ has input data denoted by the indexes in $\mathcal{M}_n$, i.e., data 
\[ \x_{n}\!\triangleq \!\x[\mathcal{M}_n]\!=\![0,\!\hdots,\!0,x_{i_{(1,n)}},0,\!\hdots,\!0,\!x_{i_{(2,n)}},\hdots,0,x_{i_{(M_n,n)}}]^{\intercal} \in \F^K, n\! \in\! [N],i_{1,n},\hdots \!i_{M_n,n}\in [K],\!  M_n\in [K]\] where $|\mathcal{M}_n|=M_n$ represents the number of bits that sender $n$ possesses and $M\triangleq \sum^{N}_{n=1} M_n$ represents the total amount of information bits that the senders store.   There are $K$ receivers $R_k,k \in [K]$, and we assume for simplicity and without loss of generality that receiver $k$ is interested in $x_k$.

\begin{Definition}(Directed Side-Information  Hypergraph)\label{side-info-graph}
We partition the edge set $\mathcal{E}$ of the directed side information hypergraph as follows:

{\color{black}
\begin{enumerate}
    \item Demand hyperedge set ($\mathcal{E}_d$): These hyperedges take the form $(\{k\},\{k,\{n\}\})\equiv (k,k,n,n)$, and represent the case where the $k$-th data block $x_k$, is found at sender $n$. 
    \item Cached data hyperedge set ($\mathcal{E}_s$): The hyperedge $(\{k\},\{k',\{n\}\})\equiv(k,k',n,n)$ exists if and only if $x_{k'}$ is stored at receiver $k\neq k'$ and at sender $n$.
    \item Coupled hyperedge set ($\mathcal{E}_c$): The hyperedge $(\{k\},\{k',\{n ,n'\}\})\equiv(k,k',n,n')$ exists if and only if $x_{k'}$ is stored at senders $n$ and $n'$ ($n\neq n'$) but not\footnote{Note that while $(k,k',n,n')=(k,k',n',n)$ must hold, it is not the case for $(k,k',n,n')=(k',k,n,n')$.} at receiver $k\neq k'$\footnote{ The pair $\{n,n'\}$ is unordered; the tuple $(k,k',n,n')$ is only a shorthand. All definitions and proofs depend solely on the unordered sender pair.}. 
\end{enumerate}}
\begin{remark}
We first quickly note that our approach naturally collapses to the single sender case, when considering only hyperedges of the form $(k,k',1,1) \in \mathcal{G}$.  We also note that, as expected, the side-information hypergraph reveals the locations of each block, across the senders and receivers. For example, the set 
\[ \mathcal{R}(k) \triangleq \{k' \in \mathcal{V} \ | \ (k,k',n,n) \in \mathcal{E}_s,  \ k \neq k' \} \]
indicates the cached data $x[\mathcal{R}(k)]$ at receiver $k$. Finally, we also see that for some $k,k'\in [K]$, if $k' \notin \mathcal{R}(k), \ k\neq k'$ and if $k' \in \mathcal{M}_n 
    \cap \mathcal{M}_{n'} \cap  \mathcal{M}_{n''}$, then $(k,k',n,n'), (k,k',n,n''),(k,k',n',n'') \in \mathcal{E}_c$. 
\end{remark}
\end{Definition}
Now that we have defined the parameters involved in the problem and the directed side-information hypergraph, we are equipped with the necessary tools to define the multi-sender index code. 
 \begin{Definition} (Multi-Sender Index Code): An index code $\mathcal{C} \in \F^K$ with side information hypergraph $\mathcal{G}$ on $K$ nodes, abbreviated as "Index code for $\mathcal{G}$", is a set of codewords in $\F^\ell$ together with:
 \begin{enumerate}
     \item $N$ encoding functions $E_n, n \in [N]$ mapping inputs $\x_n \in \{0,1\}$ to codewords in $\mathcal{C}_n \in \F^{\ell_n}$.  
     \item $K$ decoding functions $D_1,\hdots,D_K$ such that $D_k(E_1(\x_1), \hdots, E_N(\x_N), \x[\mathcal{R}(k)])=x_n$.
 \end{enumerate}
 \end{Definition}
We here consider the linear case, where $E_n, D_k, n \in [N], k \in [K]$ are linear functions of their inputs. Naturally, $\ell \triangleq \sum_{n \in [N]} \ell_n$ is the length of the used code $\mathcal{C}$. Each candidate code corresponds to a subgraph of $\mathcal{G}$, and each code is valid if the desired messages are properly recovered.
Unlike in the original single sender case though, here not all subgraphs correspond to a valid  index code, and thus we here define valid sub-hypergraphs, and use this definition to later define the hyper-minrank. 
 \begin{Definition} \label{Valid sub-hypergraph}
     (Valid sub-hypergraph) We call $\mathcal{G'}$ a valid (spanning) sub-hypergraph if and only if each node $k \in [K]$  
     appears in an odd number of demand edges of $\mathcal{G'}$.
 \end{Definition}
Towards designing a multi-sender index code, we also proceed to define a composite adjacency matrix for any side-information hypergraph and its sub-hypergraphs. We recall that $\mathcal{E}_{d},\mathcal{E}_{s},\mathcal{E}_{c},$ are respectively the demand, cached and coupled hyperedge sets.

  \begin{Definition}\label{Adjacency matrix-def} (Composite adjacency matrix of the side-information hypergraph) The composite adjacency matrix of hypergraph $\mathcal{G}$ is a $K \times KN$ matrix $\mathbf{G} \in \F^{K \times KN}$, whose rows are indexed by $k\in [K]$, whose columns are indexed by $({k,n}), k \in [K], n \in [N]$, and whose entries are as follows:
 \begin{enumerate}
     \item For any $n\in [N]$ and any $k,k'\in[K]\times[K]$, if $(k,k,n,n)\in \mathcal{E}_d$ or $(k,k',n,n)\in \mathcal{E}_s$, then $\mathbf{G}(k,(k',n)) =1$.
     \item For any $n\in[N]$ and any $k,k'\neq k$, then $\mathbf{G}(k,(k',n))) = \sum_{n' \in [N]\backslash{n}} \mathbb{1}((k,k',n,n') \in \mathcal{E}_c)$. 
    \item All other entries are set to zero.     
  \end{enumerate}
 Alternatively, for $ k, k' \in [K], n \in [N]$, $\mathbf{G}$ takes the form:
 \begin{align}
     \mathbf{G}(k,(k',n)) &= \sum_{n' \in [N]\backslash{n}} \mathbb{1}((k,k',n,n') \in \mathcal{E}_c \lor (k,k',n,n) \in \mathcal{E}_d) + \mathbb{1}((k,k',n,n) \in \mathcal{E}_s)\label{Adjacency}
 \end{align}
    \end{Definition}
    \begin{exmp}
        Suppose that a message $k'$ is stored at receiver $k$ and only at senders $n,n',n''$ (i.e., that $k' \in \mathcal{R}(k)$, and $k'\in\mathcal{M}_{n} \cap \mathcal{M}_{n'} \cap \mathcal{M}_{n''}$). After recalling that $(k,k',n,n')$, $(k,k',n,n'')$, $(k,k',n',n'')$ are all in $\mathcal{E}_c$, we see that $\mathbf{A}_{n}(k,k') = \mathbf{A}_{n'}(k,k') = \mathbf{A}_{n''}(k,k') = 1+1=0$. If on the other hand, $k' \notin \mathcal{R}(k)$ and if $k'$ is only stored at senders $n,n',n'',n'''$, then $(k,k',n,n'), (k,k',n,n''),(k,k',n,n'''),$  $(k,k',n',n''),(k,k',n',n'''), (k,k',n'',n''')$ are all in $\mathcal{E}_c$, and now $\mathbf{A}_{n}(k,k') = \mathbf{A}_{n'}(k,k') = \mathbf{A}_{n''}(k,k') = \mathbf{A}_{n'''}(k,k')=1+1+1 =1 $. Note that here $\mathbf{G} = [\mathbf{A}_1, \mathbf{A}_2,\hdots,\mathbf{A}_N]$.
    \end{exmp}
We recall from~\cite{Bar-Yossef1} that in the single-sender case, a binary matrix~$\mathbf{A}' \in \mathbb{F}^{K\times K}$ is said to \emph{fit} the side-information graph~$\mathcal{P}$ if the graph induced by~$\mathbf{A}'$ is a subgraph of~$\mathcal{P}$. 
In the multi-sender setting, this concept extends naturally by considering the hypergraph induced by~$\mathbf{A} \in \mathbb{F}^{K\times KN}$, which must form a valid (spanning) sub-hypergraph of the directed side-information hypergraph~$\mathcal{G}$. 
The following notion of fitting thus generalizes the single-sender definition, since any valid sub-hypergraph of the single-sender side-information hypergraph corresponds to a valid instance of the original (single-sender) fitting condition.

 Now we formally define the notion of fitting as follows:
 \begin{Definition} (Fitting) We say that a composite adjacency matrix  $\mathbf{A} \in \mathbf{F}^{K \times KN}$  fits the hypergraph $\mathcal{G}$ if and only if there exists a spanning hypergraph $\mathcal{A}$ whose composite adjacency matrix is $\mathbf{A}$, such that $\mathcal{A}$ is a valid sub-hypergraph of $\mathcal{G}$.
 \end{Definition}
With this definition in place, we can now --- similar to the case of the minrank function for directed graphs in \cite{Bar-Yossef1} --- define the hyper-minrank($\mathcal{G}$). We do so, as follows:
 \begin{Definition}
 \begin{align}   \hyperminrank({\mathcal{G}})\triangleq \min \{\sum^{N}_{n=1} \rank(\mathbf{A}_n)|\mathbf{A}\: \text{fits}\: \mathcal{G} \}
 \end{align}
 where $\mathbf{A}_{n}\triangleq\mathbf{A}(:,[(n-1)K+1:nK])$.
 \end{Definition}
We are now ready to provide the main result of the work, which provides optimality for any (directed) side information hypergraph, i.e. for any multi-sender index coding problem.  

\begin{theorem}\label{main-Theorem}
For any side information hypergraph, there exists a
linear Multi-Sender index code whose length equals $\hyperminrank(\mathbf{G})$. This
bound is optimal for all linear Multi-Sender index codes.
\end{theorem}
\begin{proof}
 For encoding, let $\A$ be a matrix that fits $\G$, and which satisfies sum-rank $\sum^{N}_{n=1} \rank(\mathbf{A}_n) = \hyperminrank(\mathcal{G}) =\ell$, where $\ell$ is the length of the corresponding code.  Let $\mathcal{L}_n = \{i_1,i_2,\cdots,i_{K_n}\} \subseteq [K]$ be the set of indices of the spanning rows of $\mathbf{A}_n$, where naturally $|\mathcal{L}_n|=K_n= \rank(\A_n)$, and where the rows $\a_{i_1,n},\dots, \a_{i_{K_n},n}$ of $\A_n$ span the row space of $\A_n$. The encoding function of each sender $n$ is simply $b_{k,n}= \a_{k,n}.\x_n, k \in \mathcal{L}_n$, where  $b_{k,n},  k \in \mathcal{L}_n$ is what is sent by sender $n$.

    Decoding at each receiver $R_k, k\in [K]$ proceeds as follows. Consider a sender $S_d, d\in[N]$ such that $(k,k,d,d) \in \mathcal{E}_A \subseteq \mathcal{E}_G$. First we calculate $\a_{k,d}= \sum^{}_{j \in \mathcal{L}_n }\lambda_{j,d} \a_{j,d}$  for some  $\lambda_{j,d}$\footnote{{
Such coefficients exist because the vector~$\mathbf{a}_{k,d}$ lies in the linear span of 
$\{\mathbf{a}_{j,d} : j \in \mathcal{L}_n\}$, implying that it can be expressed as a linear combination of these basis vectors.}}.  Then, receiver~$k$ evaluates 
$\mathbf{a}_{k,d}^{\top}\mathbf{x}_d = \sum_{j \in \mathcal{L}_d} \lambda_{j,d}\, b_{j,d}$ 
by linearly combining the encoded symbols received from sender~$n$, whose codeword~$\mathbf{x}_n$ represents the $K_n$-bit message segment assigned to that sender. Now consider the vector $\c_k=\a_{k,d}-\e_k$, where $\e_k$ is the $k$-the standard basis vector. Since the only nonzero entries in $\c_k$ correspond to cached-data hyperedges or coupled hyperedges, we can write $\mathbf{c}_k = \mathbf{c}'_k + \mathbf{c}''_k$ where the non-zero elements of $\mathbf{c}'_k$ correspond to cashed-data hyperedges and the non-zero elements of $\mathbf{c}''_k$ correspond to coupled hyperedges. For decoding to be successful, we have to show that $\mathbf{c}_k.\mathbf{x}_n$ can be constructed by receiver $k$ and can be subtracted from $a_{k,d} \mathbf{x}_n$ so that $\mathbf{e}_k\mathbf{x}_n = x_k$ can be obtained by receiver $k$. In order for receiver $R_k$ to do this, i.e., to cancel out the unwanted contribution $\mathbf{c}_k$, it forms $\a_{k,n}.\x_n, \forall n\in [N]$ and computes $b'_{k}\triangleq \sum_{n \in  [N]} a_{k,n}.\x_{n}$. This means that $\mathbf{b}'_k = (\mathbf{e}_k + \mathbf{c}'_k + I_{k}) \mathbf{x}_{n}$, where the support of $c'_{k}$ and $I_{k}$ are subsets of $\mathcal{R}(k)$. This in turn shows that we can %in $\mathbf{b}'_k$, 
    automatically cancel $\mathbf{c}''_{k}$, which in turn means that $b'_{k}= \sum_{j \in \mathcal{S}, \mathcal{S} \subseteq \mathcal{R}(k)}  x_j+ x_{k}$ for some subset $\mathcal{S} \subseteq \mathcal{R}(k)$, which means that receiver $k$ can successfully decode its desired message. The above is reflected in the following lemma.
    \begin{lemma}\label{mainlemma}
        We can represent $b'_{k}= \sum_{n \in [N] }a_{k,n}.\x_{n} $ as the summation $b'_{k}= \sum_{j \in \mathcal{S}, \mathcal{S} \subseteq \mathcal{R}(k)}  x_j+ x_{k}$, consisting only of messages in $\mathcal{R}(k)$ stored at receiver $R_k$, and its desired message.
    \end{lemma}
    \begin{proof}
        To prove Lemma~\ref{mainlemma}, let us first note that $\mathcal{A}$ (whose composite adjacency matrix $\mathbf{A}$ fits $\mathcal{A}$) is a valid sub-hypergraph of $\mathcal{G}$, and then directly from the definition of the composite adjacency matrix of the sub-hypergraph, we can write 
        \begin{align}
            \mathbf{a}_{k,n}.\mathbf{x}_n&= [A(k,(1,n)),\hdots,A(k,(K,n))].\mathbf{x}[\mathcal{M}_n]
            \\&=\sum_{\{j\in[K]| j\in \mathcal{M}_n\}} ( \sum_{n' \in [N]\backslash{n}}( \mathbb{1}((k,j,n,n') \in \mathcal{E}^{\mathcal{A}}_c \lor (k,j,n,n) \in \mathcal{E}^{\mathcal{A}}_d)) \nonumber\\& + \mathbb{1}((k,j,n,n) \in \mathcal{E}^{\mathcal{A}}_s) .x_{j})
        \end{align} 
        which means that
        \begin{align}
              & b_k'\overset{(a)}{=}\sum_{n\in  [N] }\a_{k,n}.\x_{n}=\\
              & \sum_{(j,n) \in [K] \times [N]} \mathbb{1}((k,j,n,n)\in \mathcal{E}^{\mathcal{A}}_s).x_{j} \nonumber\\&+  \sum_{n \in  [N] } \sum_{\{j\in[K]| j \in \mathcal{M}_n\}}  \sum_{n' \in [N]\backslash{n}} \mathbb{1}((k,j,n,n') \in \mathcal{E}^{\mathcal{A}}_c \lor (k,j,n,n) \in \mathcal{E}^{\mathcal{A}}_d).x_{j}
              \\&\overset{(b)}{=} \sum_{(j,n) \in [K] \times [N] } \mathbb{1}((k,j,n,n) \in \mathcal{E}^{\mathcal{A}}_s).x_{j} \nonumber \\&+  \sum_{\{j\in[K]\}}  \sum_{(n,n') \in [N] \times [N]\backslash{n}} \mathbb{1}((k,j,n,n') \in \mathcal{E}^{\mathcal{A}}_c).x_{j} \nonumber \\&+ \sum_{n \in  [N] } \sum_{\{j\in[K]| j \in \mathcal{M}_n\}}   \mathbb{1}((k,j,n,n) \in \mathcal{E}^{\mathcal{A}}_d).x_{j} 
               \\&\overset{(c)}{=} \sum_{(j,n) \in [K] \times [N] } \mathbb{1}((k,j,n,n) \in \mathcal{E}^{\mathcal{A}}_s) .x_{j} \nonumber\\&+  \sum_{j\in[K]}  \sum_{\{n,n'\} \subseteq [N]} (\mathbf{1}+\mathbf{1}).\mathbb{1}((k,j,n,n') \in \mathcal{E}^{\mathcal{A}}_c).x_{j}
               \nonumber\\& +(|\mathcal{E}^{\mathcal{A}}_d(k)|. \mathbf{1}).x_{k} 
               \\&\overset{(d)}{=}\sum_{j \in \mathcal{S}, \mathcal{S} \subseteq \mathcal{R}(k)}  x_j+ x_{k}
        \end{align}
        where $(a)$ follows from the fact that if $(k,j,n,n) \in \mathcal{E}_s$ holds then automatically it means that $i \in \mathcal{M}_n$ since $\mathbf{A}$ fits $\mathcal{G}$; $(b)$ follows from  the fact that $\mathcal{E}^{\mathcal{A}}_{c} \cap \mathcal{E}^{\mathcal{A}}_{d} = \emptyset$ and from the linearity of the $\Sigma$ operator as well as from the fact that if $(k,j,n,n) \in \mathcal{E}_s$ or $(k,j,n,n') \in \mathcal{E}_s$  holds then we have that $i \in \mathcal{M}_n$  since $\mathbf{A}$ fits $\mathcal{G}$; $(c)$ follows since  each coupled hyperedge corresponding to set $\{n,n'\} \subset [N]$ is considered two times in the double summation $\sum_{(n,n') \in [N] \times [N]\backslash{n}}$, and finally $(d)$ holds due to the fact that $\mathbf{A}$ is a valid sub-hypergraph of $\mathcal{G}$  and has an odd number of demand hyperedges incident to~$k$, which together with the fact that $(k,j,n,n) \in \mathcal{E}_{s}^{A}$ implies that $j \in \mathcal{R}(k)$, since the matrix~$\mathbf{A}$ fits hypergraph~$\mathcal{G}$.
  This concludes the proof of Lemma~\ref{mainlemma}.
    \end{proof}
This also \textit{concludes the achievability part of the theorem}, because from the above, we know that receiver $k$ can decode its own desired message by removing the interfering $\mathbf{b}'_{k}$ by using its own stored side-information, as well as we know that the code length is $\ell = \sum_{n \in [N]} \rank(\mathbf{A}_n)$.

For the lower bound, consider a problem (hypergraph) $\mathcal{G}$, and consider an arbitrary multi-sender linear index code $\mathcal{C}$ for this problem. Consider 
any sender $n\in[N]$, and their vector set $\mathcal{S}_n=\{\u_{1,1},\u_{1,2},\dots\,\u_{1,K_n}\}$, which instructs this sender to transmit linear combinations of its local data blocks in the form $\mathbf{u}_{i,n}^{\top}\mathbf{x}_{n}$, for all $i\in[K_n]$. 
Let $\mathcal{S}=\bigcup_{n\in[N]}\mathcal{S}_n$ be the collection of all sender sets across the senders, and let us proceed with the following lemma.
\begin{lemma}\label{Claim1}
For every $k\in[K]$, the vector $\mathbf{e}_k$ belongs to the span of $\mathcal{S}\cup\{\mathbf{e}_j: j\in\mathcal{R}(k)\}$.
\end{lemma}

\begin{proof}
Consider an arbitrary receiver $k\in[K]$, and consider any valid multi-sender
linear index code for the hypergraph~$\mathcal{G}$.  
By definition of a valid index code, receiver~$k$ must be able to recover its
desired message $x_k$ jointly from:
\begin{enumerate}
    \item the set of transmitted symbols 
    $\{\,\mathbf{u}_{i,n}^{\top}\mathbf{x}_n : \mathbf{u}_{i,n}\in\mathcal{S}_n,\ n\in[N]\,\}$
    \item its side-information $\{x_j : j\in\mathcal{R}(k)\}$.
\end{enumerate}

Let us introduce the vector space 
\[
\mathcal{V}_k \;\triangleq\;
\mathrm{span}\Big(
    \mathcal{S}
    \;\cup\;
    \{\mathbf{e}_j : j\in\mathcal{R}(k)\}
\Big)
\]
i.e., the space of all linear combinations of:
(i) all encoder vectors used by all senders,  
and (ii) the standard basis vectors associated with the messages stored at receiver $k$.

Every transmitted symbol is a linear combination of the messages:
\[
\mathbf{u}_{i,n}^{\top}\mathbf{x}_n
    \;=\;
\left(\mathbf{u}_{i,n}\right)^{\top}\mathbf{x}
\qquad (\text{with support in }\mathcal{M}_n).
\]
Since receiver~$k$ obtains every such transmitted symbol and already knows all
coordinates $\{x_j:j\in\mathcal{R}(k)\}$, linearity of the decoding operation
implies that $x_k$ must be expressible as a linear combination of these
quantities.  
Thus, there exist coefficients
$\{\alpha_{i,n}\in\mathbb{F}\}$ and $\{\beta_j\in\mathbb{F}\}$ such that
\[
x_k
    \;=\;
\sum_{n\in[N]}\;\sum_{\mathbf{u}_{i,n}\in\mathcal{S}_n}
    \alpha_{i,n}\,\big(\mathbf{u}_{i,n}^{\top}\mathbf{x}\big)
    \;+\;
\sum_{j\in\mathcal{R}(k)}\beta_j\,x_j.
\]

Rewriting the right-hand side as a single inner product with~$\mathbf{x}$ gives
\[
x_k
    \;=\;
\Bigg(
    \sum_{n}\;\sum_{\mathbf{u}_{i,n}\in\mathcal{S}_n}
        \alpha_{i,n}\mathbf{u}_{i,n}
    \;+\;
    \sum_{j\in\mathcal{R}(k)}\beta_j\mathbf{e}_j
\Bigg)^{\!\top}\mathbf{x}.
\]
Now since the left-hand side equals $\mathbf{e}_k^{\top}\mathbf{x}$, equality must
hold for each coordinate, and thus
\[
\mathbf{e}_k
    \;=\;
\sum_{n}\;\sum_{\mathbf{u}_{i,n}\in\mathcal{S}_n}
        \alpha_{i,n}\mathbf{u}_{i,n}
    \;+\;
    \sum_{j\in\mathcal{R}(k)}\beta_j\mathbf{e}_j
\in \mathcal{V}_k
\]
which proves $\mathbf{e}_k\in\mathrm{span}(\mathcal{S}\cup\{\mathbf{e}_j:j\in\mathcal{R}(k)\})$.
\end{proof}

With this lemma in place, we now know that there exist $\lambda_{i,n},\beta_{j,n}, j,i\in [K_{n}] , n \in [N]$, for which
    \begin{align}
      \e_k = \sum^{N}_{n=1} \sum^{K_n}_{i=1} \lambda_{i,n}\u_{i,n} + \sum^N_{n=1}\sum_{j \in \mathcal{R}(k) \cap \mathcal{M}_n} {\beta}_{j,n} \e_j \label{decoding-converse}
    \end{align}
and thus can conclude that there exists (sender) $d \in [N]$ for which $\u_{i,d}(k) =1, $ for\footnote{The decomposition guarantees that all the entries of $\sum^{N}_{n=1} \sum^{K_n}_{i=1} \lambda_{i,n}\u_{i,n}(\mathcal{R}_k)$ are set to zero and that all the contributions from the stored-data are accounted for in $\sum_{j \in \mathcal{R}(k)\cap \mathcal{M}_d} {\beta}_{j,n} \e_j$.} some $i \in[K_d]$. Let us now rearrange  $\eqref{decoding-converse}$ as follows:
    \begin{align}
        \e_k &= \sum^{K_d}_{i=1} \lambda_{i,d} \u_{i,d} +  \sum_{j \in \mathcal{R}(k)\cap \mathcal{M}_d} {\beta}_{j,n} \e_j +  \sum^{}_{n\in [N]\backslash{d}} \sum^{K_n}_{i=1} \lambda_{i,n}\u_{i,n} + \sum_{n \in [N]\backslash{d}} \sum_{j \in \mathcal{R}(k) \cap \mathcal{M}_n } {\beta}_{j,n} \e_j\label{decoding-criteria}
    \end{align}
  where the above holds because $\mathcal{R}(k) \subseteq \mathcal{M} = \cup_{n \in [N]} \mathcal{M}_n$, { where $\mathcal{M}$ represents the collection of all message indices that are available across the senders and demanded by the receivers.} At this point, let us construct the composite adjacency matrix of a valid sub-hypergraph of $\mathcal{G}$, by first setting 
    \begin{align}
    \a_{k,d} &= \e_k +  \sum_{j \in \mathcal{R}(k) \cap \mathcal{M}_d } {\beta}_{j,d} \e_j
    \end{align}
    where $a_{k,d}$ will be the $k$-th row of $\mathbf{A}_{d}$ of $\mathcal{A}$, the valid sub-hypergraph of $\mathcal{G}$. Let us construct this valid sub-hypergraph $\mathcal{A}$. 
 Towards this, we introduce auxiliary vectors that represent the linear reconstruction process at receiver~$k$. 
Specifically, let 
$\mathbf{b}'_{k,d} = \sum_{i=1}^{K_d} \lambda_{i,d}\,\mathbf{u}_{i,d}$ 
and 
$\mathbf{b}'_{k,n} = \sum_{i=1}^{K_n} \lambda_{i,n}\,\mathbf{u}_{i,n}$ 
denote the linear combinations of the local encoding vectors at the distributed and sender-specific layers, respectively, 
for some coefficients~$\lambda_{i,d}$ and~$\lambda_{i,n}$ chosen according to the decoding rule at receiver~$k$. 
Furthermore, define 
$\mathbf{b}''_{k,n} = \sum_{j \in \mathcal{R}(k) \cap \mathcal{M}_n} \beta_{j,n}\,\mathbf{e}_j$, 
which represents the portion of the reconstruction associated with the messages demanded by receiver~$k$ and available at sender~$n$. 
These definitions isolate the contribution of each sender to the decoding vector of receiver~$k$, 
and will be used to establish the subsequent alignment conditions.
To form $\mathcal{A}$, let us first initialize $\mathcal{E}^{\mathcal{A}}_s = \mathcal{E}^{\mathcal{A}}_d = \mathcal{E}^{\mathcal{A}}_c = \emptyset$, and proceed in the following manner.
    \begin{enumerate}
        \item For all $j,n \in [K] \times [N]$ for which $\beta_{j,n} = 1$, let us include $(k,j,n,n)$ in $ \mathcal{E}^{\mathcal{A}}_{s}$. Since $(k,j,n,n) \in \mathcal{E}^{\mathcal{G}}_{s}$ is a cached-data side-information hyperedge (cf. Definition~\ref{side-info-graph}), and since $\beta_{j,n} = 1$, we have that $j \in \mathcal{R}(k) \cap \mathcal{M}_n$, which means that $\mathcal{E}^{\mathcal{A}}_s \subseteq \mathcal{E}^{\mathcal{G}}_s$.
        \item Note that if $\mathbf{b}'_{k,d}(k) =1$ then $\sum^{}_{n\in [N]\backslash{d}} \mathbf{b}'_{k,n}(k) =0, n \in [N]\backslash\{d\}$. The reverse also holds since $\mathbf{b}'_{k,d}(k)+ \sum^{}_{n\in [N]\backslash{d}} \mathbf{b}'_{k,n}(k)=\e_k(k)$ because of \eqref{decoding-criteria}. Now let us note that the set $\{\mathbf{b}'_{k,n}(k)=1|n\in [N]\}$ \textit{has an odd number of elements}.  Let us now add the hyperedge $(k,k,n,n)$ to $ \mathcal{E}^{\mathcal{A}}_{d}$ whenever $\mathbf{b}'_{k,n}(k)=1$, and let us do so going across all $n\in[N]$. Note that $ \mathcal{E}^{\mathcal{A}}_{d} \subseteq \mathcal{E}^{\mathcal{G}}_{d}$ since  $\mathbf{b}'_{k,n}(k)=1$ if $k \in \mathbf{M}_{n}$ for some $n \in [N]$. This renders $|\mathcal{E}^{\mathcal{A}}_{d}(k)|$ odd {(which directly satisfies the validity condition from Definition~\ref{Valid sub-hypergraph}, ensuring that each receiver node $k$ participates in an odd number of demand hyperedges and hence that $\mathcal{A}$ is a valid sub-hypergraph).}
        \item For the third step, let us first define $\mathcal{S}'\triangleq [K]\backslash (\{k\} \cup (\mathcal{R}(k) \cap \mathcal{M}_d))$, and let us note that $\mathcal{S}' \subseteq\text{Supp}(\a_{k,d} + \mathbf{b}'_{k,d})$ directly due to 
        \eqref{decoding-criteria}. 
        For all $k' \in \mathcal{S}'$, if $\mathbf{b}'_{k,d}(k') =1$ then $\mathbf{b}"_{k,n}(k') =1, n \in [N]\backslash\{d\}$ since $\mathbf{b}'_{k,d}(k')+ \mathbf{b}"_{k,n}(k')=0$ (again because of~\eqref{decoding-criteria}) in which case we add $(k,k',d,n)$ in $\mathcal{E}^{\mathcal{A}}_c$.  {Note that in this case\footnote{This refers to the case where $\mathbf{b}'_{k,d}(k') = 1$ and simultaneously $\mathbf{b}''_{k,n}(k') = 1$ for some $n \in [N]\backslash\{d\}$, which by construction corresponds to the situation where receiver~$k$ does not store $x_{k'}$ while the same message $x_{k'}$ is jointly available at senders~$d$ and~$n$.} we have that $k' \in \mathcal{M}_n \cap \mathcal{M}_d$ and $k' \not\in \{k\} \cup \mathcal{R}(k)$, which directly corresponds to the defining condition of a coupled hyperedge (cf.~Definition~\ref{side-info-graph}). Consequently, we conclude that $\mathcal{E}^{\mathcal{A}}_c \subseteq \mathcal{E}^{\mathcal{G}}_c$.}

    \end{enumerate}
The hypergraph $\mathcal{A}$ is constructed by employing the above steps, for all $k\in [K]$. At this point, we verify that indeed $\mathcal{A}$ is a \textit{valid} sub-hypergraph of $\mathcal{G}$ since $|\mathcal{E}^{\mathcal{A}}_{d}(k)|$ is odd for all $k\in [K]$, and because $\mathcal{E}^{\mathcal{A}} \subseteq \mathcal{E}^{\mathcal{G}}$, as we have shown above. This tells us that the constructed composite adjacency matrix $\A$ fits $\mathcal{G}$. Furthermore, since all $\a_{k,n}$ for all $(k,n) \in[K] \times [N]$ are constructed from the sent encoded signals of sender $n$ (encoded by the vectors in  $\mathcal{S}_n$), we can also conclude that $\rank(\A_{n}) \leq K_{n}$ and thus that $\sum_{n\in [N]}\rank(\A_{n}) \leq \ell = \sum_{n\in [N]} K_{n}$. 

At this point, in conjunction with the achievability part from before, we can readily conclude that $\hyperminrank(G)$ is indeed the optimal length of the multi-sender index coding problem, and that $\hyperminrank(G) \leq \sum_{n\in [N]}\rank(\A_{n}) \leq \ell = \sum_{n\in [N]} K_{n}$. This concludes the proof of Theorem~\ref{main-Theorem}.
\end{proof}
{\begin{remark}[Embedded and classical IC as special cases]
The hyper-minrank formulation here, encapsulates the classical single-sender 
index-coding problem as well as the 
embedded multi-sender index-coding problem.  
The embedded case corresponds to the specialization $K=N$ with 
$\mathcal{M}_n = \mathcal{R}(n)$ for all $n$, under which the 4-uniform 
sender–receiver hyperedges reduce to the bipartite structure used in 
embedded index coding~\cite{Porter1,Haviv2020task,Sundar3}.  
Likewise, the classical (single-sender) index-coding problem arises as the 
special case $N=1$, in which hyperminrank coincides exactly with the 
Bar-Yossef--Birk--Jaggi minrank formulation~\cite{Bar-Yossef1}.  
\end{remark}
}
\medskip
Let us offer a small example that takes us through the encoding and decoding steps corresponding to Theorem~\ref{main-Theorem}.
\begin{exmp} Consider the binary multi-sender index coding problem, with $K=3$ receivers and $N=3$ senders, as depicted in Figure~\ref{Probelm Setting}. Let the senders store according to $\mathcal{M}_1 = \{1,2\},\mathcal{M}_2 = \{2,3\}, \mathcal{M}_3 = \{1,3\}$, and let the receivers store their side information according to $\mathcal{R}(1) = \{2\}, \mathcal{R}(2) = \{3\}, \mathcal{R}(3) = \{3\}$. Let the $3$ messages form the vector $\mathbf{x} = [x_{1},x_2,x_{3}]^{\intercal} \in \F^{3}$, which means that $\mathbf{x}_1 = [x_{1},x_{2},0]$, $\mathbf{x}_2 = [0,x_{2},x_{3}]$ and $\mathbf{x}_3 = [x_{1},0,x_{3}]$. 
    \begin{figure}
    \centering
    \includegraphics[width=0.5\linewidth]{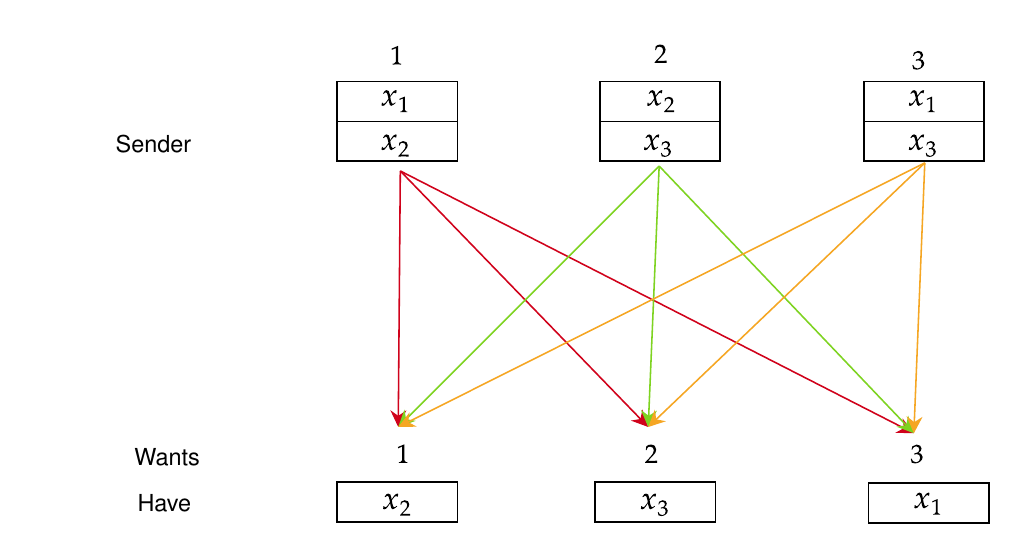}
    \vspace{-10pt}
    \caption{A multi-sender index coding problem with $K=3$, $N=3$, sender's storage $\mathcal{M}_1 = \{1,2\},\mathcal{M}_2 = \{2,3\}, \mathcal{M}_3 = \{1,3\}$ and receiver-side storage $\mathcal{R}(1) = \{2\}, \mathcal{R}(2) = \{3\}, \mathcal{R}(3) = \{3\}$.}
    \label{Probelm Setting}
\end{figure}
The first step is to create the directed side information hyper-graph, which we illustrate in  Figure~\ref{Side-information-hypergraph}, directly by following Definition~\ref{side-info-graph}. 
Then we create the composite adjacency matrix of $\mathcal{G}$, which is simply the following: 
\begin{align}
\begin{array}{r|ccc|ccc|ccc}
  & x_{1} & x_{2} & x_{3} & x_{1} & x_{2} & x_{3} & x_{1} & x_{2} & x_{3} \\ \hline
1 & \textcolor{red}{1} & \textcolor{blue}{1} & 0 & 0 & \textcolor{blue}{1} & \textcolor{green}{1}& \textcolor{red}{1} & 0 & \textcolor{green}{1} \\
2 & \textcolor{green}{1} & \textcolor{red}{1} & 0 & 0 & \textcolor{red}{1} & \textcolor{blue}{1} & \textcolor{green}{1} & 0 & \textcolor{blue}{1} \\
3 & \textcolor{blue}{1} & \textcolor{green}{1} & 0 & 0 & \textcolor{green}{1} & \textcolor{red}{1} & \textcolor{blue}{1} & 0 & \textcolor{red}{1}\\
\end{array}
\end{align}
\begin{figure}
    \centering
    \includegraphics[width=0.5\linewidth]{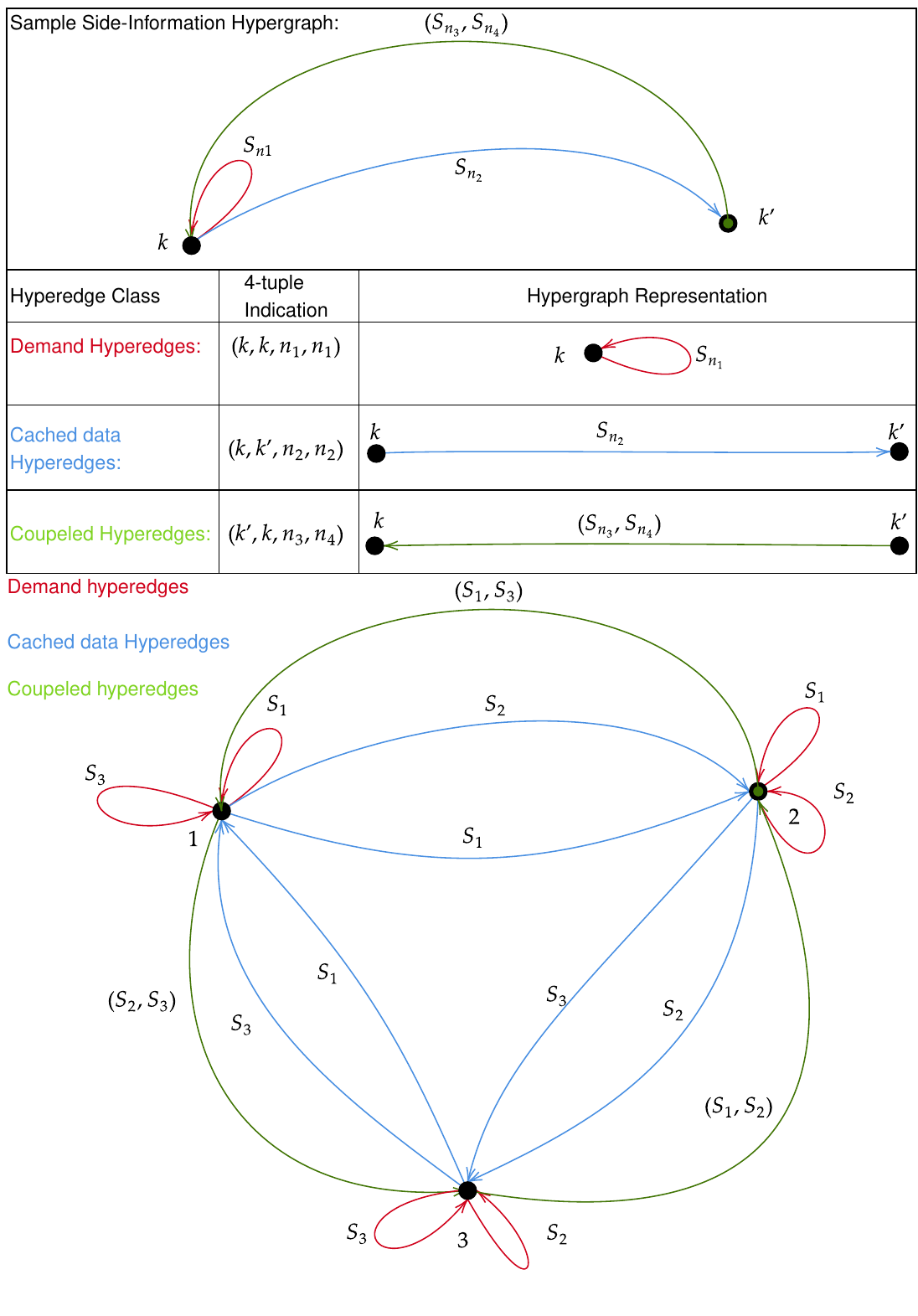}\vspace{-30pt}
    \caption{Side-information hypergraph corresponding to the problem setting in Figure~\ref{Probelm Setting}, along with its hypergraph representation guide.}
    \label{Side-information-hypergraph}
\end{figure}
Our goal now is to find a small valid sub-hypergraph, and for this example, we pick the one in Figure~\ref{valid}, whose composite adjacency matrix $\mathbf{A}$ takes the form
\begin{figure}
    \centering
    \includegraphics[width=0.5\linewidth]{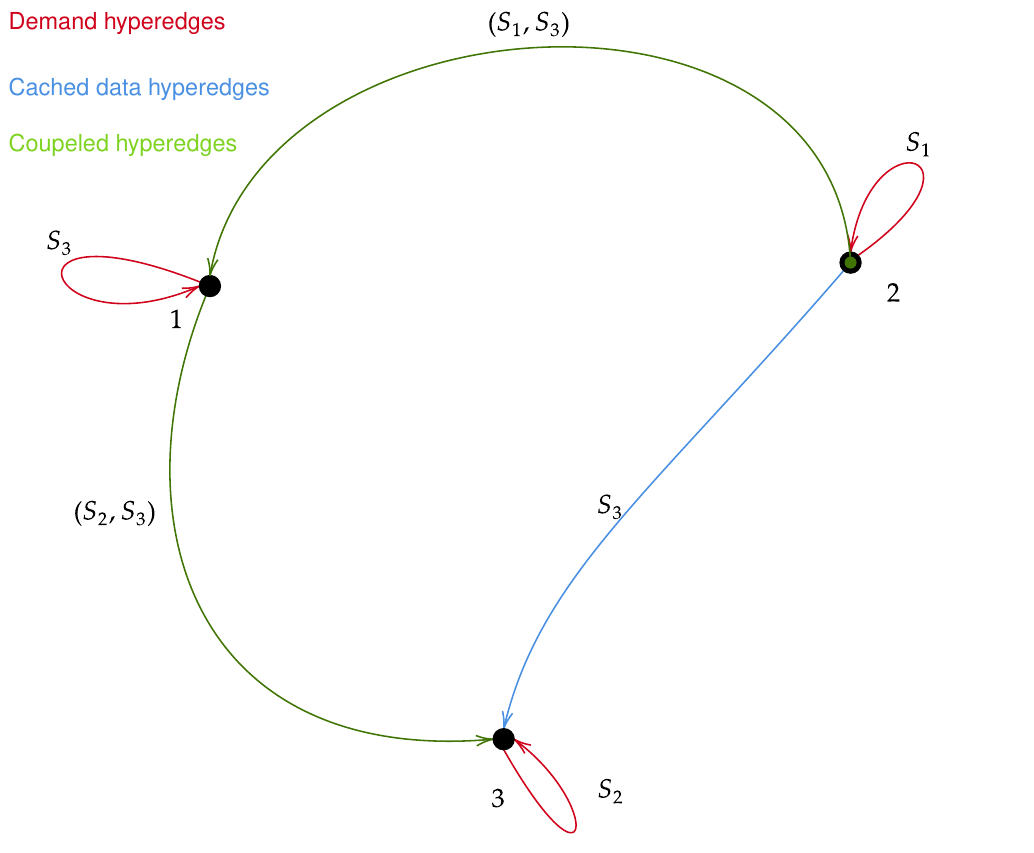}
    \vspace{-20pt}
    \caption{A valid sub-hypergraph of the side-information hypergraph presented in Figure~\ref{Side-information-hypergraph}.}
    \label{valid}
\end{figure}
\vspace{-5pt}
\begin{align}
\begin{array}{r|ccc|ccc|ccc}
  & x_{1} & x_{2} & x_{3} & x_{1} & x_{2} & x_{3} & x_{1} & x_{2} & x_{3} \\ \hline
1 & 0 & 0 & 0 & 0 & 0 & \textcolor{green}{1}& \textcolor{red}{1} & 0 & \textcolor{green}{1} \\
2 & \textcolor{green}{1} & \textcolor{red}{1} & 0 & 0 & 0 & 0 & \textcolor{green}{1} & 0 &  \textcolor{blue}{1}\\
3 & 0& 0 & 0 & 0 & 0 & \textcolor{red}{1} & 0 & 0 & 0\\
\end{array} .
\end{align}
Following the steps in the achievability part of the proof of Theorem~\ref{main-Theorem}, the above matrix tells us that sender $1$ will send ${u}_1=x_1+x_2$, sender $2$ will send ${u}_2=x_3$, and sender $3$ will send ${u}_3=x_1+x_3$. This corresponds to a total of $\sum^{3}_{i=1}\rank(\mathbf{A}_i) = 3$ transmissions. We can also verify that the code is valid, as we see that receiver $1$ can compute $\mathbf{u}_2 + \mathbf{u}_3 = x_1$ to decode $x_1$, receiver $2$ can compute ${u}_1 + u_3 = x_2 +x_3$ to decode $x_2$ after removing $x_3$ which resides in its cache, while receiver $3$ can directly use $u_2=x_3$ to get its desired symbol message.

{\vspace{0.2em}
Up to this point, the first part of the example illustrated the \emph{achievability} direction of Theorem~\ref{main-Theorem}: 
starting from a given valid sub-hypergraph $\mathcal{A}$, we explicitly constructed its composite adjacency matrix and showed how the corresponding senders’ transmissions form a valid multi-sender index code of length equal to $\sum_n \mathrm{rank}(\mathbf{A}_n)$.
We now turn to the \emph{converse} direction, aiming to show how any arbitrary multi-sender linear index code can be mapped back to a valid sub-hypergraph of $\mathcal{G}$ whose composite adjacency matrix fits the hypergraph and whose block ranks reproduce the same code length. 
In other words, this part of the example demonstrates the reverse implication of Theorem~\ref{main-Theorem} --- namely, that every valid linear code induces a valid fitting and thus cannot achieve a shorter broadcast length than $\hyperminrank(\mathcal{G})$. 
To do so, we start from a specific choice of encoding vectors at the senders and reconstruct the corresponding valid sub-hypergraph step by step, exactly mirroring the logic of the converse proof.

Now let us consider another multi-sender index code, and let us define this using the aforementioned vector sets $\mathcal{S}_n=\{\u_{1,1},\u_{1,2},\dots\,\u_{1,K_n}\}$ which will here take the form $\mathcal{S}_1 = \{\mathbf{u}_{1,1}= [ 1\: 1\: 0]\}, \mathcal{S}_2 = \{\mathbf{u}_{1,2} = [0 \:1 \:1]\}$. Recall that each vector set instructs the corresponding sender to transmit linear combinations of its local data blocks in the form $\mathbf{u}_{i,n}^{\top}\mathbf{x}_{n}$, for all $i\in[K_n]$. In our example, we now substitute this specific code structure into the general decoding decomposition of~\eqref{decoding-criteria}, obtaining:
}
 \begin{align}
        \e_1 &=  \lambda_{1,1} \u_{1,1} +   {\beta}_{2,1} \e_2 = [1 \: 1 \: 0] + [0 \: 1 \:0] = [ 1 \: 0 \: 0]\\
        \e_2 &=  \lambda_{1,2} \u_{1,2} +   {\beta}_{3,2} \e_3 = [0 \: 1 \: 1] + [0 \:0 \:1] = [0 \: 1 \: 0]\\
        \e_3 &=  \lambda_{1,2} \u_{1,2}  +\lambda_{1,1} \u_{1,1}  +   {\beta}_{1,1} \e_1 = [1 \: 1 \: 0] + [0 \: 1 \: 1] + [1 \: 0\:0] = [ 0 \: 0\: 1].
    \end{align}
The next step is to construct a valid sub-hypergraph, as shown in Figure~\ref{converse}. From this graph, we now directly construct its composite adjacency matrix $\mathbf{A'}$ which takes the form:
        \begin{align}
\begin{array}{r|ccc|ccc|ccc}
  & x_{1} & x_{2} & x_{3} & x_{1} & x_{2} & x_{3} & x_{1} & x_{2} & x_{3} \\ \hline
1 & \textcolor{red}{1} & \textcolor{blue}{1} & 0 & 0 & 0 & 0& 0 & 0 & 0 \\
2 & 0 & 0 & 0 & 0 & \textcolor{red}{1} & \textcolor{blue}{1} & 0 & 0 & 0 \\
3 & \textcolor{blue}{1}& \textcolor{green}{1} & 0 & 0 & \textcolor{green}{1} & \textcolor{red}{1} & 0 & 0 & 0\\
\end{array}.
\end{align}
Finally, we can easily see that $\sum^{3}_{n=1} \rank(\mathbf{A}'_{n}) =2$, and this matches the $\hyperminrank(G)$. 
    \begin{figure}
    \centering
    \includegraphics[width=0.5\linewidth]{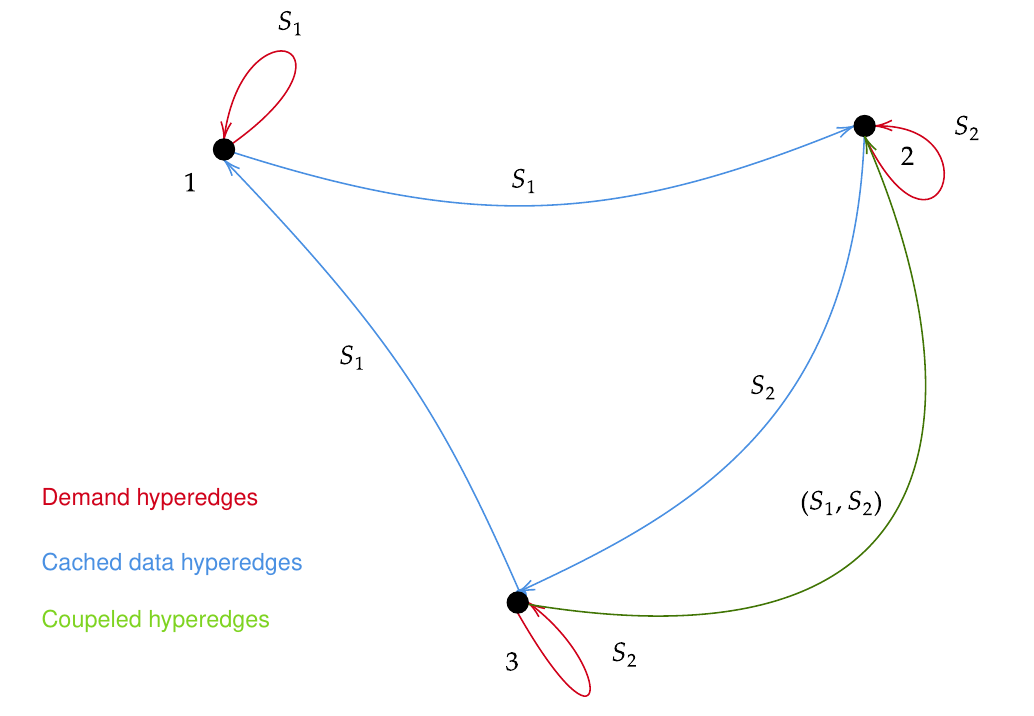}
    \vspace{-15pt}
    \caption{A valid sub-hypergraph of the side-information hypergraph presented in Figure~\ref{Side-information-hypergraph} corresponding to $\hyperminrank(\mathcal{G})$}
    \label{converse}
    \end{figure}
\end{exmp}
\section{Upper and Lower Bounds on $\hyperminrank(\mathcal{G})$}\label{sec2}
As suggested before, one of the main reasons of introducing a hypergraphic formulation for the multi-sender index coding problem is that such a formulation opens the door to leveraging the extensive body of results developed in the literature on \emph{hypergraphic bounds}. 
In particular, classical tools such as clique-cover and complement-based techniques—originally designed for hypergraphs—can now be directly employed to derive fundamental upper and lower bounds on the optimal code length. 
To do so, we begin by formalizing the notion of hypergraphic cliques and valid clique covers. We first formalize the notion of \emph{hypergraphic clique} which will capture --- in the multi-sender setting --- the same ``one-transmission'' structure that classical cliques represent in the single-sender graph model. The construction projects each 4-tuple onto its receiver and sender coordinates and enforces a uniformity condition that translates into rank-one blocks in the composite adjacency representation. We proceed with the definition. 

\begin{Definition}\label{Hypergraphic-clique}
    (Hypergraphic clique) Consider $\mathcal{C} \triangleq \{\{(k,k',n,n')|k,k' \in \mathcal{K},n,n' \in \mathcal{N}\}|\}$ where $\mathcal{K} \subseteq [K], \mathcal{N} \subseteq [N]$, as well as let $\mathcal{C}_{R} \triangleq \{(k,k')|  (k,k',n,n') \in \mathcal{C}\}$, $\mathcal{C}_{S} \triangleq  \{\{n,n'\}| \forall n \neq n':  (k,k',n,n') \in \mathcal{C}\}$, $\mathcal{C}_{S_n} \triangleq \{(k,k')|\ \: n  \in \mathcal{N} \land (k,k',n,n') \in \mathcal{C}\}$. We say that $\mathcal{C}$ is a hypergraphic clique if and only if:
    \begin{enumerate}
\item $\mathcal{C}_R$ is a clique on the vertex\footnote{A directed graph is a clique if, for any two distinct vertices~$k,k'$, it contains both arcs~$(k,k')$ and~$(k',k)$, and every vertex has a self-loop.} set~$\mathcal{K}$. We refer to this as a \emph{receiver-side clique}.

  \item $\mathcal{C}_S$ is a clique on the vertex set $\mathcal{N}$. We call this a \emph{sender-side clique}.
  \item For all $n\in[N]$ and all $\mathcal{V}\in\mathcal{C}_S$ with $n\in\mathcal{V}$, the subgraph $\mathcal{C}_{S_n}$ satisfies: for all $k,k'\in\mathcal{C}_{S_n}$, both $(k,k')$ and $(k',k)$ are in $\mathcal{C}_{S_n}$, and every $k\in\mathcal{C}_{S_n}$ has a self-loop.
    \end{enumerate}
\end{Definition}
\noindent
The next notion imposes a parity condition on the sender-side clique. This ensures that each valid clique behaves like a single effective transmission in the hypergraphic setting.

\begin{Definition}\label{Valid-Clique}(Valid Clique)
    A clique is a valid clique if and only if the degrees of each vertex in $\mathcal{C}_S$ are odd, which makes $|\mathcal{C}_S|$ an even number. 
\end{Definition}

\noindent
Having defined what constitutes a valid clique, we next introduce a simple projection operator that will be useful in describing coverage conditions for clique collections. In particular, we will need a way to identify which receivers are ``covered'' by a family of hyperedges.

\begin{Definition}[Receiver projection]\label{Receiver-projection}
For any set $\mathcal{C}$ of 4-tuples 
$(k,k',n,n')\in[K]\times[K]\times[N]\times[N]$, 
we define its \emph{receiver projection} as
\[
\mathrm{proj}_R(\mathcal{C})
\triangleq
\{\,k\in[K] \;|\; \exists\, k',n,n' \text{ such that } (k,k',n,n')\in\mathcal{C}\,\}
\]
In words, $\mathrm{proj}_R(\mathcal{C})$ collects all receivers that appear as the
first coordinate of at least one hyperedge in~$\mathcal{C}$.
For a family of hypergraph subsets $\{\mathcal{C}_1,\ldots,\mathcal{C}_m\}$,
we simply write
\[
\mathrm{proj}_R\!\Big(\bigcup_{t=1}^m\mathcal{C}_t\Big)
= \bigcup_{t=1}^m \mathrm{proj}_R(\mathcal{C}_t)
\]
\end{Definition}

\noindent
With these definitions in place, we are now ready to formalize the multi-sender analogue of the classical clique-cover bound. The following theorem establishes that, under an additional implementability condition ensuring that each clique can be served by a single sender, the number of cliques in such a cover provides a tight upper bound on the optimal broadcast length.

\begin{theorem}[Clique-cover upper bound with implementability]\label{Theorem-upper}
Let $\mathcal{G}$ be the directed side-information hypergraph of a multi-sender index coding instance with composite adjacency matrix $\mathbf{G}$.
Assume there exist $m$ \emph{valid hypergraphic cliques} $\mathcal{C}_1,\dots,\mathcal{C}_m$ such that:
\begin{enumerate}
    \item (Receiver coverage) The receiver projection of their union covers all receivers:
    \[
      \mathrm{proj}_R\!\Big(\,\bigcup_{t=1}^m \mathcal{C}_t\Big)
      \;=\;[K]
    \]
    \item (Validity) $\mathcal{H}\triangleq\bigcup_{t=1}^m\mathcal{C}_t$ is a valid sub-hypergraph of $\mathcal{G}$ (Definition~\ref{Valid sub-hypergraph}).
    \item (Implementability) For each clique $\mathcal{C}_j$ there exists a sender $n_j\in[N]$ such that all demanded messages indexed by the receiver set $\mathcal{K}_j$ of $\mathcal{C}_j$ are stored at $n_j$, i.e., $\mathcal{K}_j\subseteq \mathcal{M}_{n_j}$.
\end{enumerate}
Then the optimal linear broadcast length satisfies
\[
\hyperminrank(\mathcal{G}) \;\le\; m
\]
\end{theorem}
The proof proceeds by showing that each implementable valid clique
can be served by a single sender using one rank-one transmission,
and that the union of all such cliques yields a fitting composite
adjacency matrix of total rank at most $m$. We recall that all sums are over the binary field $\F_2$.
\begin{proof}
We show that the above assumptions yield a linear multi-sender index code with exactly $m$ transmissions.

\medskip
\noindent\textbf{Step 1:}
Fix a valid hypergraphic clique $\mathcal{C}_j$ with receiver set $\mathcal{K}_j$ and sender projection $\mathcal{C}_{j_S}$.
By Definition~\ref{Hypergraphic-clique} and the composite adjacency rule \eqref{Adjacency}, for every $(k,k')\in\mathcal{C}_{j_R}$ and every sender $n$ appearing in $\mathcal{C}_{j_S}$, we have
\[
\mathbf{G}(k,(k',n)) \;=\; 1
\]
Indeed we see that:
(i) for $k=k'$ we have a \emph{demand} hyperedge $(k,k,n,n)\in\mathcal{E}_d$ (Definition~\ref{side-info-graph});
(ii) for $k\neq k'$ and $(k,k',n,n)\in\mathcal{C}_j$ we have a cached-data hyperedge $(k,k',n,n)\in\mathcal{E}_s$;
(iii) for $k\neq k'$ and $(k,k',n,n')\in\mathcal{C}_j$ with $n\neq n'$ we have a coupled hyperedge in $\mathcal{E}_c$, and then the parity condition of a \emph{valid} clique (Definition~\ref{Valid-Clique}) guarantees that the coupled/demand sum in \eqref{Adjacency} contributes a $1$.
Hence the submatrix of $\mathbf{G}$ with rows $\mathcal{K}_j$ and columns $\{(k',n):k'\in\mathcal{K}_j,\ n\in\mathcal{C}_{j_S}\}$ is an all-one matrix, and thus each nonempty sender block corresponding to $\mathcal{C}_j$ has rank one.

\medskip
\noindent\textbf{Step 2:}
First recall that by the \emph{implementability} assumption above, there exists a sender $n_j$ such that \(\mathcal{K}_j\subseteq \mathcal{M}_{n_j}\).
Proceed to construct a block matrix $\mathbf{A}_{j}$ that \emph{keeps only} the sender block at $n_j$, i.e., first construct
\[
\mathbf{A}_{j,n_j} \;\text{has rows indexed by }\mathcal{K}_j\text{ and columns } \{(k',n_j):k'\in\mathcal{K}_j\}
\]
and set $\mathbf{A}_{j,n}=\mathbf{0}$ for all $n\neq n_j$.
By Step~1, $\mathbf{A}_{j,n_j}$ has rank one (its entries over $\mathcal{K}_j\times\mathcal{K}_j$ are all ones), and thus
\[
\sum_{n=1}^N \rank(\mathbf{A}_{j,n}) \;=\; 1
\]

\medskip
\noindent\textbf{Step 3:}
Let $\mathbf{A}_{\mathcal{H}}\triangleq \sum_{j=1}^m \mathbf{A}_{j}$. % be the blockwise sum over senders. 
Because $\mathcal{H}=\cup_{j=1}^m\mathcal{C}_j$ is a valid sub-hypergraph of $\mathcal{G}$, the composite adjacency matrix $\mathbf{A}_{\mathcal{H}}$ \emph{fits} $\mathcal{G}$ (Definition~\ref{Valid sub-hypergraph} and Definition~\ref{Adjacency matrix-def}). 
By construction,
\[
\sum_{n=1}^N \rank(\mathbf{A}_{\mathcal{H},n})
\;\le\;\sum_{j=1}^m \sum_{n=1}^N \rank(\mathbf{A}_{j,n})
\;=\; m
\]

\medskip
\noindent\textbf{Step 4:}
 By Theorem~\ref{main-Theorem} (achievability–converse equivalence for hyper-minrank), any fitting composite adjacency matrix yields a linear index code whose length equals the sum of sender-block ranks. 
Applying this to $\mathbf{A}_{\mathcal{H}}$ gives a code of length at most $m$, and thus
\[
\hyperminrank(\mathcal{G}) \;\le\; m
\]
\end{proof}
\noindent
The following corollary is a direct consequence of Theorem~\ref{Theorem-upper}.
It restates the result in terms of the minimal number of valid hypergraphic cliques
needed to cover the entire directed side-information hypergraph.

   \begin{corollary}[Hypergraphic clique-cover bound]
The hyper-minrank of the directed side-information hypergraph $\mathcal{G}$ of any multi-sender index-coding problem is upper-bounded by its hypergraphic clique-cover number, i.e.,
by the minimum number of valid hypergraphic cliques whose union covers $\mathcal{G}$ and forms a valid sub-hypergraph of $\mathcal{G}$.
\end{corollary}
\begin{proof}
    The proof is direct from Theorem~\ref{Theorem-upper}.
\end{proof}
\medskip
\noindent
We now turn to the \emph{lower} bound. Let us recall that in the single-sender case, clique numbers of the complement graph offer a lower-bound for the minrank. 
In our multi-sender case though, the presence of coupled edges complicates the concept of a complement graph, and thus, in this context, we here define a complement $\bar{\mathcal{G}}$ that removes coupled edges by collapsing their effect into single-sender constraints, in a way that matches our adjacency rule.
\begin{Definition}\label{Complement}(Complement of the side-information hypergraph)
Let $\mathcal{G}$ be the side-information hypergraph of a multi-sender index-coding  problem. Then $\Bar{\mathcal{G}}$ is said to be the complement of $\mathcal{G}$ if and only if all the following hold:
\begin{enumerate}
    \item If $(k,k,n,n) \in \mathcal{G}$ for some  $k \in [K], n \in [N]$, then $(k,k,n,n) \in \bar{\mathcal{G}}$.
     \item If $(k,k,n,n) \notin \mathcal{G}$ for some  $k \in [K], n \in [N]$, then $(k,k,n,n) \notin \bar{\mathcal{G}}$.
    \item If $(k,k',n,n) \in \mathcal{G}  $ for some $k,k'\in [K], n' \in [N]$, then $(k,k',n,n) \notin \bar{\mathcal{G}}$.
     \item If $ (k,k',n,n') \in \mathcal{G} $ for some $k,k'\in [K], n,n' \in [N], n\neq n'$, then $(k,k',n,n) \notin \bar{\mathcal{G}}$ and  $(k,k',n',n') \notin \bar{\mathcal{G}}$.
    \item If $(k,k',n,n) \notin \mathcal{G} $ for some $k,k'\in [K], n \in [N]$, then $(k,k',n,n) \in \bar{\mathcal{G}}$.
    \item If $ (k,k',n,n') \notin \mathcal{G} $ for some $k,k'\in [K], n,n' \in [N],n \neq n'$, then $(k,k',n,n) \in \bar{\mathcal{G}}$ and $(k,k',n',n') \in \bar{\mathcal{G}}$. 
\end{enumerate}
\noindent
To clarify, conditions (3) and (4) restrict how $\bar{\mathcal{G}}$ may reuse demands that are already represented in $\mathcal{G}$. 
Condition~(3) forbids $\bar{\mathcal{G}}$ from containing any hyperedge with the same demand–sender tuple as an existing hyperedge in $\mathcal{G}$. 
Condition~(4) focuses on coupled hyperedges, i.e., hyperedges of the form $(k,k',n,n')$ with $n\neq n'$ that involve two senders simultaneously. 
If such a coupled hyperedge is present in $\mathcal{G}$, then $\bar{\mathcal{G}}$ is not allowed to contain any hyperedge that serves the same demand $(k,k')$ from either sender $n$ or sender $n'$ alone. 
Together, these restrictions ensure that $\bar{\mathcal{G}}$ does not introduce additional sender-specific realizations of demands that are already encoded in $\mathcal{G}$.
\end{Definition}

Let us now define the directed graphs 
\begin{align}
\bar{\mathcal{G}_{n}} &\triangleq \{(k,k')| (k,k',n,n) \in  \bar{\mathcal{G}}\} , n \in[N], \\
\mathcal{C}_n &\triangleq \{\mathcal{C}\:|\:\mathcal{C} \: \text{\:is a directed graph  clique}, \mathcal{C} \subseteq \bar{\mathcal{G}}_{n}\}
\end{align}
based on which, we have the following theorem. 
    \begin{theorem}\label{Theorem-lower}
        Let $\mathcal{C} \in \cup_{n \in [N]} \mathcal{C}_n$ be the largest clique such that, for all $n \in [N]$, either $\mathcal{C} \cap \bar{\mathcal{G}}_n = \mathcal{C}$ holds or $(k,k) \notin \mathcal{C} \cap \bar{\mathcal{G}}_n, \forall k \in \mathcal{V}(\mathcal{C})  $ holds. Then $|\mathcal{V}(\mathcal{C})| \leq \hyperminrank({\mathcal{G}})$.
    \end{theorem}
    \begin{proof}
    Suppose that $\mathcal{C} \in \cup_{n \in [N]} \mathcal{C}_n$, and 
        without loss of generality, also suppose that $\mathcal{V}(\mathcal{C}) =\{1,2,\hdots,\tilde{K}\}$, $\tilde{K} \in [K]$. We will show that any multi-sender index-coding scheme involves at least $|\mathcal{V}(\mathcal{C})|$  transmissions.  Towards this, let us consider $\bar{\mathbf{G}} =[\bar{\mathbf{G}}_1, \bar{\mathbf{G}}_2,\hdots, \bar{\mathbf{G}}_N] \in \F^{K \times KN}, \bar{\mathbf{G}}_{n} \in \F^{K \times K}, \forall n \in [N] $ to be the composite adjacency matrix of $\bar{\mathcal{G}}$, and similarly let us consider $\mathbf{G}$ to be the composite adjacency matrix of $\mathcal{G}$. 
        Considering: (i) the fact that for some $n \in [N]$ and $k,k' \in [K]$ with $k \neq k'$, we have $\bar{\mathbf{G}}_n(k,k') = 1$ (by the definition of the composite adjacency matrix of the side-information hypergraph; cf.~Definition~\ref{Adjacency matrix-def}), and (ii) the fact that $\bar{\mathcal{G}}$ does not contain any coupled-edge (cf.~Definition~\ref{Complement}), we can conclude that $(k,k',n,n) \in \bar{\mathcal{G}}$.
        Since $(k,k',n,n) \in \bar{\mathcal{G}}$ (cf.~Definition~\ref{Complement}, items 5) and 6)), we can conclude that $(k,k',n,n') \notin {\mathcal{G}}, \forall n' \in [N]$, and thus that $\mathbf{G}_n(k,k')=0$.
{
Also consider that for some $n \in [N]$ and $k' \in [K]$, we have $\bar{\mathbf{G}}_n(k,k') = 1$. 
By applying the definition of the composite adjacency matrix of the side-information hypergraph (cf.~Definition~\ref{Adjacency matrix-def}) 
and by noting, from Definition~\ref{Complement}, that $\bar{\mathcal{G}}$ contains no coupled edges, 
we deduce that $(k,k,n,n) \in \bar{\mathcal{G}}$. 
Since $(k,k,n,n) \in \bar{\mathcal{G}}$, we can conclude (cf.~Definition~\ref{Complement}, items~1) and~2)) that $(k,k,n,n) \in \mathcal{G}$ and thus that $\mathbf{G}_n(k,k) = 1$.}

        Now, since $\mathcal{C}$ either satisfies $\mathcal{C} \cap \bar{\mathcal{G}}_n = \mathcal{C}$, $n \in [N]$, or satisfies $(k,k) \notin \mathcal{C} \cap \bar{\mathcal{G}}_n, \forall k \in \mathcal{V}(\mathcal{C})$, we can conclude that submatrix $\bar{\mathbf{G}}_{n}([\tilde{K}], [\tilde{K}]) = \mathbf{1}_{\tilde{K} \times \tilde{K}}$ or that $\bar{\mathbf{G}}_{n}([\tilde{K}], [\tilde{K}]) = \mathbf{M} \in \F^{{\tilde{K}} \times {\tilde{K}}}$ where $\mathbf{M}(i,i) =0, \forall i \in [\tilde{K}]$.
        Recall that $\mathcal{C}$ is a receiver-side clique with vertex set 
$\mathcal{V}(\mathcal{C}) = \{1,2,\ldots,\tilde{K}\}$, where 
$\tilde{K} = |\mathcal{V}(\mathcal{C})|$ denotes the number of receivers in this clique.  
By the arguments above, for every sender $n \in [N]$ the composite adjacency matrix 
$\mathbf{G}_n$ must satisfy a consistent structural constraint, discussed below, on the 
$\tilde{K} \times \tilde{K}$ submatrix corresponding to the receivers in $\mathcal{C}$.  

In particular, using the conclusions derived from the complement hypergraph 
and the absence of coupled-edges in $\bar{\mathcal{G}}$, we deduce that the restriction
of $\mathbf{G}_n$ to the rows and columns indexed by $\mathcal{V}(\mathcal{C})$ 
must take one of the following two forms:
\[
\mathbf{G}_n([\tilde{K}], [\tilde{K}]) 
   = \mathbf{I}_{\tilde{K} \times \tilde{K}}
\]
where $\mathbf{I}_{\tilde{K} \times \tilde{K}}$ is the $\tilde{K} \times \tilde{K}$ identity 
matrix over $\mathbb{F}$, or
\[
\mathbf{G}_n([\tilde{K}], [\tilde{K}]) 
   = \tilde{\mathbf{M}} \in \mathbb{F}^{\tilde{K} \times \tilde{K}},
\qquad 
\tilde{\mathbf{M}}(i,i) = 0 \; \text{for all } i \in [\tilde{K}]
\]
in which case the diagonal entries corresponding to the receivers in the clique are forced 
to be zero.  

These two possibilities exhaust all admissible structures for the submatrix induced by 
the clique $\mathcal{C}$ in $\mathbf{G}_n$, and they follow directly from the 
hypergraph-complement constraints described above.By referring to Definitions~\ref{side-info-graph} and~\ref{Adjacency matrix-def}, 
we can now conclude the following.  
For all $k,k' \in [K]$ with $k \neq k'$ and $k' \notin \mathcal{R}(k)$, 
and for every sender 
$n \in \{\,n' \mid \mathcal{C} \subseteq \bar{\mathcal{G}}_{n'}\,\}$, 
it must hold that $[K] \subseteq \mathcal{M}_n$.  
Similarly, for all senders 
$n \in \{\,n' \mid (k,k) \notin \mathcal{C} \cap \bar{\mathcal{G}}_{n'},\ \forall k \in \mathcal{V}(\mathcal{C})\,\}$, 
we must have $k \notin \mathcal{M}_n$ for every $k \in [\tilde{K}]$.  
Therefore, any composite adjacency matrix of a valid sub-hypergraph 
$\mathbf{A} \in \mathbb{F}^{K \times KN}$, 
written as $[\mathbf{A}_1,\ldots,\mathbf{A}_N]$, 
that fits $\mathcal{G}$ must satisfy:

    \begin{align}\label{argument}
        \exists \: n \in \{n'|\mathcal{C} \cap \bar{\mathcal{G}}_n' = \mathcal{C}\}:\mathbf{A}_n([\tilde{K}],[\tilde{K}]) = \mathbf{I}_{\tilde{K} \times \tilde{K}}
    \end{align}
    and thus we conclude that $\tilde{K} \leq \sum_{n \in [N]}{\rank(\mathbf{A}_{n})}$, which in turn means that $\tilde{K} \leq \hyperminrank({\mathbf{G}})$ since \eqref{argument} holds for all fitting $\mathbf{A}$.          
    \end{proof}
    
\noindent
Taken together, the clique-cover upper bound and the complementary clique-number lower bound constitute a hypergraphic analogue to the classical graph theoretic bounds for minrank, now tailored to the multi-sender setting and its coupled-edge structure.

\section{Algorithms}\label{sec3}
In this section, we present the algorithm that optimally resolves each instance of the multi-sender index coding problem, and then we analyze its computational complexity and search space, and compare these with those in~\cite{kim3} and~\cite{li2018multi}. Additionally, within the context of embedded index coding—where each sender also acts as a receiver—we benchmark our algorithm against those in~\cite{Porter1}.
 \subsection{Optimal Algorithm for Multi-Sender Index Coding}

Let us first briefly outline the steps of the algorithm for computing the exact value of $\hyperminrank(\mathcal{G})$ for a given multi-sender side-information hypergraph $\mathcal{G}$.

\emph{Step 1 (Construct all valid candidates consistent with parity):}
For each receiver~$k$, we enumerate all admissible selections of demand and cached edges, and for every $k'\neq k$, all even-parity combinations of coupled edges. 
These parity constraints locally enforce the \emph{valid sub-hypergraph} condition (Definition~\ref{Valid sub-hypergraph}) at each receiver, ensuring that every matrix generated corresponds to some valid sub-hypergraph $\mathcal{G}'\subseteq\mathcal{G}$. 
This structured enumeration dramatically prunes the search space compared with a naïve exhaustive search.

\emph{Step 2 (Evaluate cost):}
For each candidate composite adjacency matrix $\{\mathbf{A}_n\}_{n=1}^N$ obtained in Step~1, we compute 
\[
R=\sum_{n=1}^N\rank(\mathbf{A}_n)
\]
which represents the broadcast length of the corresponding linear index code (Theorem~\ref{main-Theorem}).

\emph{Step 3 (Select the optimal solution):}
Track the smallest value of~$R$ over all candidates. 
The minimal value found equals the desired $\hyperminrank(\mathcal{G})$.

\begin{algorithm}[H]\label{Algo}
\caption{Computation of Hyper-Minrank}
\KwIn{Directed Side-Information Hypergraph of Multi-Sender Index Coding Problem $\{\mathcal{G}\}$}
\KwOut{hyper-minrank}

\textbf{Initialization:} Set $\text{hyper-minrank} \gets K$.

\BlankLine
\textbf{Step 1:}   \For{$k \in [K]$}{
set $\mathcal{N}_{c,k}(k') \gets \{n,n' \mid (k,k',n,n') \in \mathcal{E}_{c,k}\}, \forall k' \in [K]\backslash\{k\}$

\ForEach{collection of subsets $\mathcal{A}_d  \subseteq \mathcal{E}_{d,k}$, $\mathcal{A}_s \subseteq \mathcal{E}_{s,k}$  and  $\mathcal{A}_c(k') \subseteq \mathcal{N}_{c,k}(k'),k'\in [K] \backslash\{k\}$ such that $|\mathcal{A}_d|\equiv 1 \pmod 2$ and $|\mathcal{A}_c(k')|\equiv 0 \pmod 2$}{
     Form matrices $\mathbf{A}_n \in \mathbb{F}^{K \times K}$, $n \in [N]$. 

    \For{$(k,k') \in [K]^2 \times [N]$}{
        \eIf{$(k,k',n,n) \in \mathcal{A}_d \cup \mathcal{A}_s$ or $n\in \mathcal{A}_c(k')$ }{
            set $A_n(k,k') \gets 1$ \;
        }{
            set $A_n(k,k') \gets 0$ \;
        }
    }
}}
\textbf{Step 2:} Compute $R \gets \sum_{n=1}^N \operatorname{rank}(\mathbf{A}_n)$.

\textbf{Step 3:} If $R < \text{hyper-minrank}$ then set $\text{hyper-minrank} \gets R$.

\textbf{Return} $\text{hyper-minrank}$.
\end{algorithm}

\medskip
A direct exhaustive search over all spanning sub-hypergraphs of $\mathcal{G}$ would be computationally infeasible. 
The following lemma establishes that \emph{Step~1} enumerates exactly the set of composite adjacency matrices corresponding to all valid sub-hypergraphs—no more and no less. 
Hence, every candidate evaluated in Step~2 is feasible, and no feasible candidate is missed, proving the completeness of the search under the imposed parity constraints.
    \begin{lemma}
         All possible adjacency matrices of valid sub-hypergraphs $\mathbf{A}_\mathbf{G}', \mathcal{G}' \subseteq \mathcal{G}$ are formed and checked by Step 1 of Algorithm~\ref{Algo}. 
    \end{lemma}
\begin{proof}
Let $\mathcal{G}' \subseteq \mathcal{G}$ denote any valid sub-hypergraph, and let its composite adjacency matrix be 
$\mathbf{A}_{\mathcal{G}'}=[\mathbf{A}_1,\mathbf{A}_2,\ldots,\mathbf{A}_N]$. 
We denote by $\mathcal{E}$ and $\mathcal{E}'$ the edge sets of $\mathcal{G}$ and $\mathcal{G}'$, respectively. 

Recall that, in Step~1 of Algorithm~\ref{Algo}, for each receiver~$k$ the algorithm enumerates subsets
\[
\mathcal{A}_{d,k} \subseteq \mathcal{E}_{d,k}, \qquad 
\mathcal{A}_{s,k} \subseteq \mathcal{E}_{s,k}, \qquad 
\mathcal{A}_{c,k}(k') \subseteq \mathcal{N}_{c,k}(k'),~\forall k'\neq k
\]
where
\[
\mathcal{N}_{c,k}(k') \triangleq 
\{\,n\in[N]~|~\exists\,n'\in[N]\setminus\{n\}:(k,k',n,n')\in\mathcal{E}_{c,k}\,\}
\]
These sets represent, respectively, the demand, cached, and coupled hyperedges incident to receiver~$k$.  
Step~1 enforces the parity constraints $|\mathcal{A}_{d,k}|\equiv1\pmod2$ and $|\mathcal{A}_{c,k}(k')|\equiv0\pmod2$, 
ensuring that each generated adjacency matrix corresponds to a \emph{valid sub-hypergraph} (Definition~\ref{Valid sub-hypergraph}).

\smallskip
To prove completeness, we show that for every valid $\mathcal{G}'\subseteq\mathcal{G}$, there exists a corresponding collection 
$\{\mathcal{A}_{d,k},\mathcal{A}_{s,k},\mathcal{A}_{c,k}(k')\}$, defined exactly as in the algorithm, that reconstructs $\mathbf{A}_{\mathcal{G}'}$.

\paragraph*{Step 1 (Mapping from $\mathcal{G}'$ to the algorithmic subsets)}
By validity of $\mathcal{G}'$, each receiver~$k$ has an odd number of demand hyperedges, i.e., 
$\mathcal{E}'_{d,k}\subseteq\mathcal{E}_{d,k}$ with $|\mathcal{E}'_{d,k}|\equiv1\pmod2$. 
We therefore set
\[
\mathcal{A}_{d,k} = \mathcal{E}'_{d,k}
\]
which directly satisfies the required parity condition and matches the algorithm’s construction.  
Similarly, since $\mathcal{G}'\subseteq\mathcal{G}$, we set
\[
\mathcal{A}_{s,k} = \mathcal{E}'_{s,k}
\]
For coupled hyperedges, we also define
\[
\mathcal{A}_{c,k}(k') 
= 
\big\{
n\in\mathcal{N}_{c,k}(k')~\big|~
\sum_{n'\in[N]\setminus\{n\}}
\mathbb{1}\{(k,k',n,n')\in\mathcal{E}'_{c,k}\}=1
\big\}
\]
In words, $\mathcal{A}_{c,k}(k')$ consists of all senders~$n$ such that, in $\mathcal{G}'$, there exists exactly one other sender~$n'$ forming a coupled hyperedge $(k,k',n,n')$.  
Because $\mathcal{E}'_{c,k}\subseteq\mathcal{E}_{c,k}$, it follows immediately that $\mathcal{A}_{c,k}(k')\subseteq\mathcal{N}_{c,k}(k')$.

\paragraph*{Step 2 (Parity of $\mathcal{A}_{c,k}(k')$)}
Each coupled hyperedge $(k,k',n,n')$ contributes two directed edges—$(k,k',n,n')$ and $(k,k',n',n)$—thus over $\mathbb{F}_2$ the total number of coupled edges incident on receiver $k$ is even, i.e.,
\begin{align}\label{eq1}
\sum_{n\in[N]}\sum_{n'\in[N]\setminus\{n\}}
\mathbb{1}\{(k,k',n,n')\in\mathcal{E}'_{c,k}\}
&=
\sum_{\{n,n'\}\subseteq[N]}
(1+1)\,\mathbb{1}\{(k,k',n,n')\in\mathcal{E}'_{c,k}\}
=0
\end{align}
On the other hand, the left-hand side of~\eqref{eq1} can be written equivalently as
\begin{align}\label{eq2}
\sum_{n\in[N]}\sum_{n'\in[N]\setminus\{n\}}
\mathbb{1}\{(k,k',n,n')\in\mathcal{E}'_{c,k}\}
=|\mathcal{A}_{c,k}(k')|
\end{align}
Comparing~\eqref{eq1} and~\eqref{eq2} shows that $|\mathcal{A}_{c,k}(k')|$ must be even, which matches the parity constraint enforced in Step~1.

\paragraph*{Step 3 (Recovering adjacency entries)}
If $(k,k',n,n)\in\mathcal{A}_{d,k}\cup\mathcal{A}_{s,k}$, 
then by Definition~\ref{Adjacency matrix-def}, $\mathbf{A}_n(k,k')=1$.  Now consider the case where $n\in\mathcal{A}_{c,k}(k')$.  
From Definition~\ref{Adjacency matrix-def},
\begin{align}
\mathbf{A}_n(k,k') 
&=\sum_{n'\in[N]\setminus\{n\}}
\mathbb{1}\!\big((k,k',n,n')\in\mathcal{E}'_c
\lor (k,k',n,n)\in\mathcal{E}'_d\big)
+\mathbb{1}\!\big((k,k',n,n)\in\mathcal{E}'_s\big)\\
&\overset{(a)}{=}
\sum_{n'\in[N]\setminus\{n\}}
\mathbb{1}\!\big((k,k',n,n')\in\mathcal{E}'_c\big)
\overset{(b)}{=}1
\end{align}
In the above, (a) holds because having $n\in\mathcal{A}_{c,k}(k')$ implies that there exists some $n'\in[N]$ such that $(k,k',n,n')\in\mathcal{E}'_{c,k}$,  
and hence $(k,k',n,n)$ cannot also appear in $\mathcal{E}'_{d,k}\cup\mathcal{E}'_{s,k}$, 
since these edge classes are disjoint.  Furthermore, (b) follows directly from the definition of $\mathcal{A}_{c,k}(k')$, which ensures that for each such $n$, exactly one corresponding $n'$ contributes to the sum, yielding a value of~1.

\paragraph*{Step 4 (Converse case)}
Conversely, if $(k,k',n,n)\notin\mathcal{A}_{d,k}\cup\mathcal{A}_{s,k}$ 
and $n\notin\mathcal{A}_{c,k}(k')$, then no corresponding edge contributes.  
Substituting into~\eqref{Adjacency} and noting that
\[
(k,k',n,n)\notin\mathcal{E}'_{d,k}\cup\mathcal{E}'_{s,k}
\quad\text{and}\quad
\sum_{n'\in[N]\setminus\{n\}}
\mathbb{1}\{(k,k',n,n')\in\mathcal{E}'_c\}=0
\]
we obtain $\mathbf{A}_n(k,k')=0$.

\smallskip
Hence, every valid sub-hypergraph $\mathcal{G}'$ produces corresponding subsets 
$\mathcal{A}_{d,k},\mathcal{A}_{s,k},\mathcal{A}_{c,k}(k')$ 
exactly as enumerated in Step~1 of Algorithm~\ref{Algo}, 
and conversely, each such combination yields a valid $\mathbf{A}_{\mathcal{G}'}$. 
Therefore, Step~1 enumerates \emph{all and only} the adjacency matrices of valid sub-hypergraphs of~$\mathcal{G}$.
\end{proof}
  To compute the complexity of  Algorithm~\ref{Algo}, we first have to calculate the search space. This  equates to the number of odd-cardinality subsets of $\mathcal{E}_{d,k}, k \in [K]$,  subsets of $\mathcal{E}_{s,k}, \forall k \in [K]$  and even-cardinality subsets $\mathcal{N}_{c,k}(k'), \forall k \in[K], k' \in [K] \backslash{\{k'\}}$. Let us denote the size of this search space as $S$. Now let us note that:
    \begin{align}
        S &\overset{(a)}{=} \Pi_{k \in [K]} (2^{(|\mathcal{E}_{d,k}|-1 + |\mathcal{E}_{s,k}|)} \Pi_{k' \in [K]\backslash{\{k\}}} 2^{(|\mathcal{N}_{c,k}(k')| -1)}) \\&\overset{(b)}{=} 2^{\sum_{k \in [K]}( |\mathcal{M}(k)| -1 + |\mathcal{R}(k)| + \sum_{k' \in [K] \backslash{ \mathcal{R}(k) \cup \{k\}}}(|\mathcal{M}(k')| -1))}
    \end{align}
where $\mathcal{M}(k) \triangleq \{n | k \in \mathcal{M}_{n}\}$, where (a) is true because of Step 1 of Algorithm~\ref{Algo}, and where (b) results from the definitions of $\mathcal{E}_d, \mathcal{E}_{s}$ and $\mathcal{E}_c$. Step 2 requires the computation of the rank of $N$ matrices of size $ K\times K$ in $\F$, thus the complexity takes the form:
\begin{align} \label{eq:C_1}
     C_1 =2^{\sum_{k \in [K]}( |\mathcal{M}(k)| -1 + |\mathcal{R}(k)| + \sum_{k' \in [K] \backslash{ \mathcal{R}(k) \cup \{k\}}}(|\mathcal{M}(k')| -1))} N K^3
\end{align}
{Here, the factor $NK^3$ reflects the fact that for each sender $n\in[N]$, the algorithm must iterate over all ordered triples $(k,k',k'')\in[K]^3$
when building and checking candidate cliques in the composite adjacency structure.
}
\subsection{Comparison}
To compare our exact algorithm with the optimal method of \cite{Kim1}, we rewrite the latter’s order–complexity, in terms of our notation, as 
\begin{align}
C_2
=\;2^{\sum_{k \in [K]}\!\Big(\sum_{n \in [N]}\big(\,|\mathcal{R}(k)\cap\mathcal{M}_n| \;+\; |([K]\setminus \mathcal{R}(k))\cap \mathcal{M}_n \cap \mathcal{M}_c|\,\big)
\;-\; |([K]\setminus \mathcal{R}(k))\cap \mathcal{M}_c| \Big)} \; NK^3
\label{eq:C2}
\end{align}
where we consider
\[
\mathcal{M}_c \;\triangleq\; \big\{\, m\in[K] \;:\; \exists\, n\neq n'\in[N]\ \text{with } m\in\mathcal{M}_n \text{ and } m\in\mathcal{M}_{n'} \,\big\}
\]
to be the \emph{replicated–message set}, that is, the set of all message indices available at more than one sender (i.e., messages stored redundantly across different $\mathcal{M}_n$’s). 
Here, the common polynomial factor $NK^3$ reflects rank computations and feasibility checks, as in our case.

To relate the search–space sizes of the two \emph{optimal} algorithms, we use the following lemma, which upper–bounds the exponent of~\eqref{eq:C2} (corresponding to the optimal exhaustive algorithm of~\cite{Kim1}) by that of our method.

\begin{lemma}[Complexity comparison]\label{lem:C1_le_C2_min}
With $C_1$ defined in~\eqref{eq:C_1} and $C_2$ defined in~\eqref{eq:C2} (from~\cite{Kim1}), the two complexities satisfy
\[
C_1 \;\le\; C_2
\]
with equality if and only if $|\mathcal{M}(m)|=1$ for every $m\in\mathcal{R}(k)$ and every $k\in[K]$, i.e., with equality only in the extreme case where no side-information symbol is replicated across senders.
\end{lemma}

\begin{proof}
As both $C_1$ and $C_2$ share the term $NK^3$, we will simply compare the two exponential terms
$E_i = \log C_i / NK^3, \ i=1,2$. For the first case, let $E_1=\sum_{k\in[K]}E_1(k)$ where, for each $k$, 
We directly have 
\[
E_1(k) \;=\; (|\mathcal{M}(k)|-1) \;+\; |\mathcal{R}(k)|
\;+\!\!\sum_{m\in[K]\setminus(\mathcal{R}(k)\cup\{k\})}\!\!(|\mathcal{M}(m)|-1)
\]
Since $[K]\setminus\mathcal{R}(k)=\{k\}\cup([K]\setminus(\mathcal{R}(k)\cup\{k\}))$, 
we can merge the isolated $m=k$ term into the summation, yielding
\begin{equation}\label{eq:E1-merge}
E_1(k)
=|\mathcal{R}(k)|
+\sum_{m\in[K]\setminus\mathcal{R}(k)} (|\mathcal{M}(m)|-1)
\end{equation}
If $m\notin\mathcal{M}_c$ (i.e., $m$ is stored at exactly one sender), then $|\mathcal{M}(m)|=1$, 
and thus $(|\mathcal{M}(m)|-1)=0$. 
We can thus conclude that
\begin{equation}\label{eq:E1-final}
E_1(k)
=|\mathcal{R}(k)|
+\sum_{m\in([K]\setminus\mathcal{R}(k))\cap\mathcal{M}_c} (|\mathcal{M}(m)|-1)
\end{equation}
where again recall that $E_1=\sum_{k\in[K]}E_1(k).$

For the second case, in order to simplify $E_2=\sum_{k\in[K]}E_2(k)$, we use the standard double-counting identity: 
for any $\mathcal{A}\subseteq[K]$,
\begin{equation}\label{eq:DC}
\sum_{n\in[N]}\big|\mathcal{A}\cap\mathcal{M}_n\big|
=\sum_{m\in\mathcal{A}}|\mathcal{M}(m)|.
\end{equation}
Fix $k$. 
Applying~\eqref{eq:DC} with $\mathcal{A}=\mathcal{R}(k)$ 
and again with $\mathcal{A}=([K]\setminus\mathcal{R}(k))\cap\mathcal{M}_c$, we obtain
\begin{align}
E_2(k)
&=\sum_{m\in\mathcal{R}(k)}|\mathcal{M}(m)|
+\sum_{m\in([K]\setminus\mathcal{R}(k))\cap\mathcal{M}_c}|\mathcal{M}(m)|
-\big|([K]\setminus\mathcal{R}(k))\cap\mathcal{M}_c\big|\notag\\
&=\sum_{m\in\mathcal{R}(k)}|\mathcal{M}(m)|
+\sum_{m\in([K]\setminus\mathcal{R}(k))\cap\mathcal{M}_c}\big(|\mathcal{M}(m)|-1\big)
\label{eq:E2-final}
\end{align}
Then, subtracting~\eqref{eq:E1-final} from~\eqref{eq:E2-final}, we get
\[
E_2(k)-E_1(k)
=\Big(\sum_{m\in\mathcal{R}(k)}|\mathcal{M}(m)|\Big)-|\mathcal{R}(k)|
=\sum_{m\in\mathcal{R}(k)}\big(|\mathcal{M}(m)|-1\big)\ \ge\ 0
\]
because $|\mathcal{M}(m)|\ge1$ for all $m$. {
Summing over $k$ gives
\[
E_2 - E_1 = \sum_{k\in[K]}\sum_{m\in\mathcal{R}(k)}\big(|\mathcal{M}(m)| - 1\big) \ge 0
\]
which implies that $E_1 \le E_2$. 
Because the exponential mapping $x \mapsto 2^x$ is strictly increasing on $\mathbb{R}$, 
inequalities between exponents are preserved when exponentiated; that is, 
if $E_1 \le E_2$, then $2^{E_1} \le 2^{E_2}$. 
Hence,
\begin{align}\label{exponComparison}
C_1 = N K^3 \cdot 2^{E_1} \;\le\; N K^3 \cdot 2^{E_2} = C_2
\end{align}}

Equality holds if and only if every term $|\mathcal{M}(m)|-1$ in the final expression vanishes, 
that is, if $|\mathcal{M}(m)|=1$ for all $m\in\mathcal{R}(k)$ and all $k\in[K]$.
\end{proof}

\begin{remark}[Intuition on computational gains]
To understand $C_1$, let us recall from~\eqref{eq:E1-final} that $E_1(k) =|\mathcal{R}(k)| +\sum_{m\in([K]\setminus\mathcal{R}(k))\cap\mathcal{M}_c} (|\mathcal{M}(m)|-1)$, which reveals how $C_1$ first captures a search space complexity of $|\mathcal{R}(k)|$ for each receiver~$k$ (one unit of cost per side-information symbol) and then considers the additional costs stemming from replicated symbols that lie \emph{outside} the receiver’s known set $\mathcal{R}(k)$—that is, for every message $m$ not in $\mathcal{R}(k)$ but stored at multiple senders, as captured by the sum of $(|\mathcal{M}(m)|-1)$ over all $m\in([K]\setminus\mathcal{R}(k))\cap\mathcal{M}_c$. Intuitively, each such replicated message outside $\mathcal{R}(k)$ introduces one extra degree of combinatorial freedom in the search.
{ In contrast, $C_2$ accounts not only for these \emph{external} replications but also for replications that occur \emph{among} the side-information sets $\{\mathcal{R}(k)\}$ themselves. 
This internal contribution appears through the additional term $\sum_{m\in\mathcal{R}(k)}|\mathcal{M}(m)|$, which increases the exponent whenever side-information symbols are duplicated across multiple receivers or senders.}

The difference between the two counts—the term $\sum_{k}\sum_{m\in\mathcal{R}(k)}(|\mathcal{M}(m)|-1)$—is therefore always nonnegative. 
It quantifies the “extra charge’’ that $C_2$ pays for replicated symbols that are already known to the receivers, explaining why $C_2$ is never smaller than $C_1$ and why equality holds only when no side-information symbol is replicated across senders.
\end{remark}

\begin{exmp}[Complexity comparison]\label{ex:toy-example}
Let $K{=}3$, $N{=}2$, $\mathcal{M}_1=\{1,2\}$, $\mathcal{M}_2=\{2,3\}$, so $|\mathcal{M}(1)|{=}1$, $|\mathcal{M}(2)|{=}2$, $|\mathcal{M}(3)|{=}1$ and $\mathcal{M}_c=\{2\}$. 
Take $\mathcal{R}(1)=\{2\}$, $\mathcal{R}(2)=\{1\}$, $\mathcal{R}(3)=\{2\}$.
From \eqref{eq:E1-final}, $E_1=1+2+1=4$. From \eqref{eq:E2-final}, we see that $E_2=2+2+2=6$, which allows us to conclude that $C_1=2^{4}\,NK^3 < 2^{6}\,NK^3=C_2$.
Indeed, the inequality is strict whenever some $m\in\mathcal{R}(k)$ is replicated across senders (i.e., $|\mathcal{M}(m)|\ge 2$).
\end{exmp}

Let us now consider the approximate solution given in \cite{li2018multi}, for which, in terms of our notations, the order of complexity of its algorithm is:
\begin{align}
    C_3 = 2^{\sum_{n \in [N]} \frac{|\mathcal{M}_n|^2 + |\mathcal{M}_n|}{2} } (\sum_{n \in [N]} |\mathcal{M}_n|^2 + \sum_{(k,n) \in [K] \times [N]}|[K] \backslash{\mathcal{R}(k)}|^2 |\mathcal{M}_n|)\label{eq:C3}
\end{align}
An additional important aspect is that Algorithm~1 presented here computes the \emph{exact} optimal broadcast rate for the instances under study, whereas the method of~\cite{li2018multi} attains exactness only when exhaustive enumeration is feasible and otherwise relies on the LT-CMAR algorithm. 
Although often effective empirically, the LT-CMAR algorithm does not always return the true optimum—discrepancies are reported on larger instances (e.g., with five or more receivers and/or in the presence of substantial amounts of replications of data across the senders), yielding a nonzero gap between the heuristically evaluated scheme (and performance) and the corresponding exact broadcast rate. Moreover, while~\cite{li2018multi} does not provide a formal approximation guarantee for the heuristic, it includes a general lower bound, whose tightness, though, is unknown.
From a computational standpoint, the search space in~\cite{li2018multi} scales as in~\eqref{eq:C3}, i.e., with an exponential term that is \emph{quadratic} in the per-sender loads $\{|\mathcal{M}_n|\}_{n\in[N]}$, whereas the exponent of our exact search $C_1$ grows only \emph{linearly} with the replication surplus and side-information sizes. 
Consequently, under moderate replication and limited side information, our exact algorithm is provably less complex than~\eqref{eq:C3}, as established in the next lemma. In the following, we have \begin{align} \label{eq:C3b}
C_3 \;=\; 2^{\sum_{n \in [N]} \frac{|\mathcal{M}_n|^2 + |\mathcal{M}_n|}{2} }
\Big(\,\sum_{n \in [N]} |\mathcal{M}_n|^2 \;+\; \sum_{(k,n) \in [K]\times[N]}\! \big|[K]\setminus \mathcal{R}(k)\big|^{2}\, |\mathcal{M}_n| \Big)
\end{align}
{ representing the exhaustive-search cost of the LT--CMAR suboptimal algorithm (cf.~\cite{li2018multi}), 
while we recall that $C_1 = 2^{E_1}\, N K^3$ represents the complexity of Algorithm~1.}

\begin{lemma}\label{lem:C1_vs_C3}
Under the bounded replication constraint $\sum_{n} |\mathcal{M}_n| \le (1+\delta)K, \ \delta\in[0,1)$, and under a bounded side information constraint $|\mathcal{R}(k)| \le r_0$ for all $k$, then, for sufficiently large $K$, having 
\begin{equation}\label{eq:threshold}
\frac{r_0}{K} + \delta \;\le\; \frac{(1+\delta)^2}{2N}
\end{equation}
guarantees that $C_1 \le C_3$. 
In particular, with \emph{no replication} ($\delta=0$), it suffices that $K \ge 2N r_0$. 
Moreover, whenever the inequality in~\eqref{eq:threshold} is \emph{strict}, 
the ratio $\tfrac{C_3}{C_1}$ grows exponentially in~$K$, yielding an unbounded computational gain in favor of Algorithm~1.
\end{lemma}

\begin{proof}
We compare exponents and ignore lower-order polynomial factors. 
From the derivation of $C_1$ (see~\eqref{eq:C_1}), we first see that
\[
E_1 \;=\; \sum_{k}|\mathcal{R}(k)| \;+\; \sum_{k}\sum_{m\notin \mathcal{R}(k)}\big(|\mathcal{M}(m)|-1\big)
\;\le\; K r_0 \;+\; K\!\sum_{m}\big(|\mathcal{M}(m)|-1\big)
\]
By double counting $\sum_m |\mathcal{M}(m)|=\sum_n |\mathcal{M}_n|$, we then have
$\sum_m (|\mathcal{M}(m)|-1)=\sum_n |\mathcal{M}_n| - K \le \delta K$, which gives
\begin{equation}\label{eq:E1_bound}
E_1 \;\le\; K r_0 + \delta K^2.
\end{equation}
Then, focusing on $C_3$, we see that its exponential term carries an   exponent
\begin{equation}\label{eq:E3_bound}
E_3 \;=\; \sum_{n}\frac{|\mathcal{M}_n|^2+|\mathcal{M}_n|}{2}
\;\ge\; \frac{1}{2}\sum_{n}|\mathcal{M}_n|^2
\;\ge\; \frac{1}{2N}\Big(\sum_{n}|\mathcal{M}_n|\Big)^2
\;\ge\; \frac{(1+\delta)^2}{2N}\,K^2
\end{equation}
{where the above follows from applying the Cauchy--Schwarz inequality to~\eqref{eq:E3_bound} under the 
 \emph{low-replication regime of }$\sum_{n\in[N]}|\mathcal{M}_n|\le(1+\delta)K$, which implies that the total message load grows 
almost linearly\footnote{
The omitted additive term 
$+\tfrac{1}{2}\sum_{n}|\mathcal{M}_n|$ 
is nonnegative and thus its inclusion would only increase the right-hand side of~\eqref{eq:E3_bound}, 
yielding an even larger (tighter) lower bound on $E_3$. 
Omitting it, therefore, simplifies the expression without weakening the inequality. 
Since the subsequent comparison $E_1\le E_3$ is preserved under such monotone strengthening, the proof remains valid.
} with $K$.}{As we have seen before, focusing on the complexity exponents suffices for our comparison; A sufficient condition for $C_1 \le C_3$ in the asymptotic regime—namely, as the number 
of receivers $K$ grows large while the number of senders $N$ and the message–availability 
parameters remain fixed—is that the exponential factor in our method does not exceed the 
exponential factor of LT--CMAR, i.e., that $E_1 \le E_3$.  
This reflects the fact that the dominant contribution to both $C_1$ and $C_3$ comes from 
their exponential dependence on $K$, whereas all polynomial factors (such as $NK^3$, $K^2$, 
and similar terms) become negligible compared to the exponential growth.
} 
Using the bounds in~\eqref{eq:E1_bound}--\eqref{eq:E3_bound}, this inequality is guaranteed whenever
\[
\frac{r_0}{K} + \delta \;\le\; \frac{(1+\delta)^2}{2N}
\]
In words, this condition states that the normalized side-information density ($r_0/K$) together with the replication surplus ($\delta$) must not exceed the scaled quadratic load term $(1+\delta)^2/(2N)$.  
In the replication-free case ($\delta=0$), the inequality simplifies to
\[
K \;\ge\; 2N r_0
\]
which means that when the number of receivers is at least twice the total ``side-information budget'' across all senders, our algorithm is guaranteed to be exponentially more efficient than LT--CMAR.

Finally, if the inequality in~\eqref{eq:threshold} is strict, then $E_3-E_1=\Omega(K^2)$, so 
\[
\frac{C_3}{C_1}
\;=\;2^{E_3-E_1}
\;\ge\;2^{\,\Omega(K^2)}
\]
which validates the statement that under these assumptions, Algorithm~1 achieves an \emph{exponentially unbounded computational gain} over the LT--CMAR approach as $K$ grows.
\end{proof}
\begin{remark}[Comparison with LT--CMAR]
The LT--CMAR~\cite{li2018multi} search complexity $C_3$ grows as $2^{\sum_n (|\M_n|^2+|\M_n|)/2}$, thus \emph{quadratically} in the per-sender loads $|\M_n|$. 
In contrast, our $C_1$ grows as $2^{E_1}$, where $E_1$ depends only \emph{linearly} on the replication surplus $\sum_m (|\M(m)|-1)$ and on the side-information sizes $\{|\mathcal{R}(k)|\}$. 
Therefore, when replication is modest ($\delta$ small) and each receiver knows only a few packets ($r_0\!\ll\!K$), 
the quadratic exponent of $C_3$ eventually outpaces the linear exponent of $C_1$, 
making Algorithm~1 \emph{asymptotically preferable for larger problem instances}.
\end{remark}

\paragraph*{Embedded / decentralized single–unicast comparison} Let us consider now the decentralized embedded single–unicast setting where each sender is also a receiver, where naturally $K=N$, and where each node's stored data serves also as its side information, corresponding to the case of $\mathcal{M}_n=\mathcal{R}(n)$ for all $n\in[N]$. 
Let $d_m \triangleq |\mathcal{M}(m)|$ denote the replication degree of message $m\in[K]$, and let 
$S \triangleq \sum_{n\in[N]}|\mathcal{M}_n|=\sum_{m\in[K]} d_m$.
In this setting, the approximate algorithm of \emph{Embedded Index Coding}~\cite{Porter1} involves a search space that is exponential in $S$ with polynomial-time checks (see \cite[Sec.~5]{Porter1}, for their heuristic procedures and complexity discussion), so in our notation its order complexity is
\[
C_{EIC}\;\triangleq \;2^{\,S}\,\mathrm{poly}(K)\;\approx\;2^{\sum_{n}|\mathcal{M}_n|}\,K^3
\]
This complexity\footnote{In the above, the term $K^3$ reflects standard rank computations and feasibility checks; any comparable polynomial factor would suffice for the purpose of our comparisons.} is generally lower than that of Algorithm~1, which, however, yields the \emph{exact} optimal broadcast rate—unlike the algorithm proposed in~\cite{Porter1}. This is captured in the lemma below. 
\begin{lemma}[Embedded case]
\label{lem:EIC_exact_vs_approx}
In the embedded IC setting with $K=N$ and $\mathcal{M}_n=\mathcal{R}(n)$ for all $n$, the complexity $C_1 \;\approx\; 2^{\,E_1}\,K^3$ of Algorithm~1 entails an exponent
\(
E_1 \;=\; S \;+\; \sum_{m=1}^{K} (d_m-1)\,(K-d_m),
\)
while the complexity the approximate embedded scheme satisfies $C_{EIC}\approx 2^{\,S}K^3$. Consequently,
\(
C_{EIC}\ \le\ C_1,
\)
with equality if and only if $d_m\in\{1,K\}$ for every $m\in[K]$, corresponding to the case where each message is stored at exactly one node or at all nodes.
\end{lemma}
{
\begin{proof}
In the embedded setting, every message $m$ is stored at exactly $d_m$ senders, where
$d_m \triangleq |\{n : m\in\mathcal{M}_n\}|$, and the total number of stored
message instances is $S=\sum_{k=1}^{K}|\mathcal{R}(k)|.$ In Algorithm~1, each time a receiver $k$ does \emph{not} cache message $m$
(i.e., for each of the $(K-d_m)$ such receivers), the algorithm must explore
which sender among the $d_m$ holders might serve $m$ to receiver $k$.  
Only $d_m-1$ of these possibilities create nontrivial search branches, since
one holder corresponds to the “direct” uncoded option.  
Thus message $m$ contributes an exponential branching factor of
\[
(d_m-1)(K-d_m).
\]

Summing this over all messages, and noting that the baseline cost $S$
accounts for the $S$ message-request incidences, the total exponential
exponent of Algorithm~1 is
\[
E_1 \;=\; S + \sum_{m=1}^{K}(d_m-1)(K-d_m)
\]
leading to complexity
\[
C_1 \approx 2^{E_1}K^3
\]
where the factor $K^3$ captures the polynomial-time adjacency and
consistency checks per branch.

The approximate embedded scheme of~\cite{li2018multi} does not branch on
replication and therefore enjoys an exponent $S$, and thus 
\[
C_{EIC} \approx 2^{S}K^3
\]

Since
\[
E_1-S=\sum_{m}(d_m-1)(K-d_m)\ge 0
\]
we obtain $C_{EIC}\le C_1$, with equality if and only if every term
$(d_m-1)(K-d_m)$ is zero.  
This occurs precisely when $d_m\in\{1,K\}$ for all $m$, i.e., when each
message is stored either at a single sender or at all senders. This proves the lemma.
\end{proof}
}
\begin{remark}[Complexity Comparison for EIC] The approximate embedded-index-coding complexity $C_{EIC}$ scales exponentially only with the total number of cached symbols $S$. In contrast, the exact complexity $C_1$ includes an additional replication penalty $\sum_m (d_m-1)(K-d_m)$, which becomes strictly positive whenever a message is stored at more than one but fewer than $K$ senders. This term captures the extra branching required to account for partially replicated messages in the exact search.
In contrast, our exact $C_1$ incurs the same baseline cost $S$ plus an additional replication–versus–coverage penalty 
$\sum_m (d_m-1)(K-d_m)$, which becomes positive whenever a message is stored at multiple, but not all, senders. 
Intuitively, this term reflects the extra search needed to verify all feasible combinations of partially replicated messages across different nodes. 
This is precisely the additional computational effort required to certify optimality.{\footnote{As a sanity check to verify that $E_1 \ge S$, observe that 
$E_1 - S = \sum_m (d_m - 1)(K - d_m)$, 
which is termwise nonnegative since $d_m \in [1,K]$. 
Equivalently, using $E_1 = (K + 2)S - \sum_m d_m^2 - K^2$ together with 
$S = \sum_m d_m$ yields the same inequality.}}
\end{remark}

\section{Conclusions}\label{sec4}
This work extends the classical minrank framework, from the single sender to the multi sender index coding setting. A first key ingredient to our approach is the introduction of a directed side-information \emph{hypergraph} with hyperedges that encode demands and side information, while also capturing the newly identified \emph{coupled} inter-sender interactions. A second key ingredient is the introduction of a tailored adjacency-and-fitting set of rules, defined so that every valid sub-hypergraph induces a valid linear code and, conversely, every linear multi-sender index code induces a valid fitting. 
Although a $4$-regular hypergraphic formalism may appear restrictive at first glance, our approach leads to a matching achievability-converse statement that says that the optimal scalar linear broadcast length equals $\hyperminrank(\mathcal{G})$. As an additional consequence of this approach, we have also established multi-sender analogues of Haemers-type graph-theoretic bounds by defining appropriate cliques and complements on valid sub-hypergraphs, yielding clique-cover upper bounds and complementary clique-number lower bounds for $\hyperminrank(\mathcal{G})$, thereby extending classical graph-theoretic tools to the distributed setting. 
These bounds can be useful; in traditional single-sender index coding, most concise and insightful characterizations of performance have historically stemmed from \emph{graph-theoretic} constructions—such as chromatic, independence, or clique-based arguments—rather than from direct algebraic or minrank analyses alone. 
Our results thus bridge these two viewpoints by embedding graph-theoretic intuition within a rigorous minrank-type formulation applicable to the multi-sender domain.

\paragraph*{Explicit algorithmic design} On the algorithmic side, we presented an \emph{exact} procedure to compute $\hyperminrank(\mathcal{G})$, explicitly characterized its search space and complexity, and revealed the computational gains over the state of art~\cite{Kim1}. Additional analysis also revealed a performance-vs-complexity tradeoff with the LT--CMAR approach of~\cite{li2018multi} for the embedded index-coding setting~\cite{Porter1}. While the worst-case complexity remains exponential—as is typical for exact combinatorial solvers—many practical instances are significantly easier due to exploitable structure (such as sparse side information, sender/receiver symmetries, and bounded replication). Moreover, the clique-cover upper bounds and complementary clique-number lower bounds allow the algorithm to prune large portions of the search space early, thereby accelerating practical performance.

\paragraph*{Potential applications} The here proposed framework has immediate applications in systems 
where different nodes or senders store distinct, possibly overlapping, subsets of a global message library, and where communication takes place over bottleneck (broadcast) links. 
Such systems relate to, for example, multi-sender coded caching\cite{7580630} where cache placement (of uncoded data) can map naturally to distributed index-coding instances, and where now the newly introduced coupled hyperedges can capture cross-sender interference cancellation during the coded caching delivery phase. A similar connection can also appear with coded distributed computing\cite{li2017fundamental,8437333,10458969} (entailing MapReduce clusters with overlapping data shards), or with distributed storage and repair where helpers act as senders.  Strong connections also arise with distributed computing over multi-sender and multi-receiver settings~\cite{9614153,e26060448,namboodiri2025fundamentallimitsdistributedcomputing,10146295,10947203}, where multi-sender index coding provides a unifying analytical framework—much like classical index coding has long done for single-sender systems.
Direct links also appear in other high-impact systems naturally modeled by multi-sender index coding, especially in those cases where data placement at receivers and transmitters is not designer-controlled. Such systems include IoT platforms, vehicular or edge networks with partial replication across nodes, and even modern satellite systems, where multiple satellites or gateways hold overlapping content and jointly serve ground terminals with storage capabilities. This includes GEO/LEO deployments with cached user data, multi-gateway architectures with partial replication, and payloads with limited inter-satellite exchange; all settings where the current hyper-minrank approach can help design more efficient downlink transmissions with reduced spectrum.

Beyond quantifying efficiency, our approach also serves as a design tool: by capturing the role of redundancy and interference, our coupled-edge approach reveals how content should be placed across transmitters and receivers to enable effective joint transmission and cancellation.  In the end, as modern systems feature increasingly structured data and multiple storage/compute nodes serving many receivers, the multi-sender index-coding viewpoint captures two key realities: communication over bottleneck links and data replicated across transmitters and receivers, often in unplanned, non-structured ways. In all such cases, $\hyperminrank$ provides a principled target for delivery-rate optimization, while our graph-theoretic bounds offer insightful performance guarantees when exact computation is unnecessary.

\paragraph*{Future directions}
Natural extensions of this work include:
\begin{itemize}
    \item Moving beyond $\mathbb{F}_2$ to larger fields and to vector (blocklength-$L$) linear codes.
    \item Developing randomized models for \emph{typical-case analysis} under random placements and random side-information hypergraphs, including the study of concentration bounds and phase transitions as replication or sparsity parameters vary.
    \item Sharpening relaxations via LP/SDP\footnote{Linear Programming and Semidefinite Programming}-based {hypergraph analogues of Lovász-type bounds} to obtain efficiently computable \emph{certificates}, in the form of provable upper or lower bounds that can be verified in polynomial time.
    \item Co-designing {placement and delivery} for multi-sender coded caching within the present hypergraph-based framework. The current paper focuses on the delivery phase, while joint design remains an open direction.
    \item Exploring non-linear schemes, as well as robustness to noisy links and adversarial perturbations.
    \item Studying scalable solvers through symmetry reductions, problem decomposition, and rank-update–based branch-and-bound search strategies.
\end{itemize}

In the end, the proposed hypergraph formulation, the exact $\hyperminrank$ characterization, the associated clique-based bounds, and the algorithmic design can jointly form a useful toolkit that can be applied in a variety of distributed systems of high potential.

\appendices
\section{Illustrative Example}\label{Example}
Let us present a simple example that can help the reader better understand the various steps, from the construction of the side-information hypergraph, to the computation of $\hyperminrank(\mathcal{G})$, to the use of the two bounds in Theorem~\ref{Theorem-upper} and Theorem~\ref{Theorem-lower}. 

    \begin{figure}
    \centering
\includegraphics[width=0.6\linewidth]{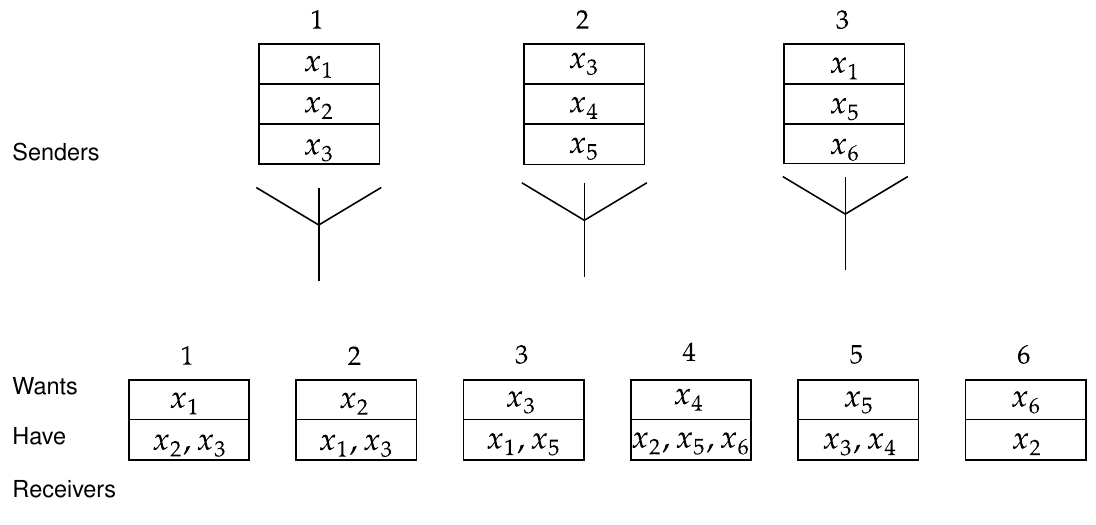}
\vspace{-20pt}
    \caption{Appendix~\ref{Example} (Illustrative Example): A $K=6$ receiver and $N=3$ sender, multi-sender index coding problem. }
    \label{BigExample1}
    \end{figure}
\begin{figure}
    \centering
\includegraphics[width=0.5\linewidth]{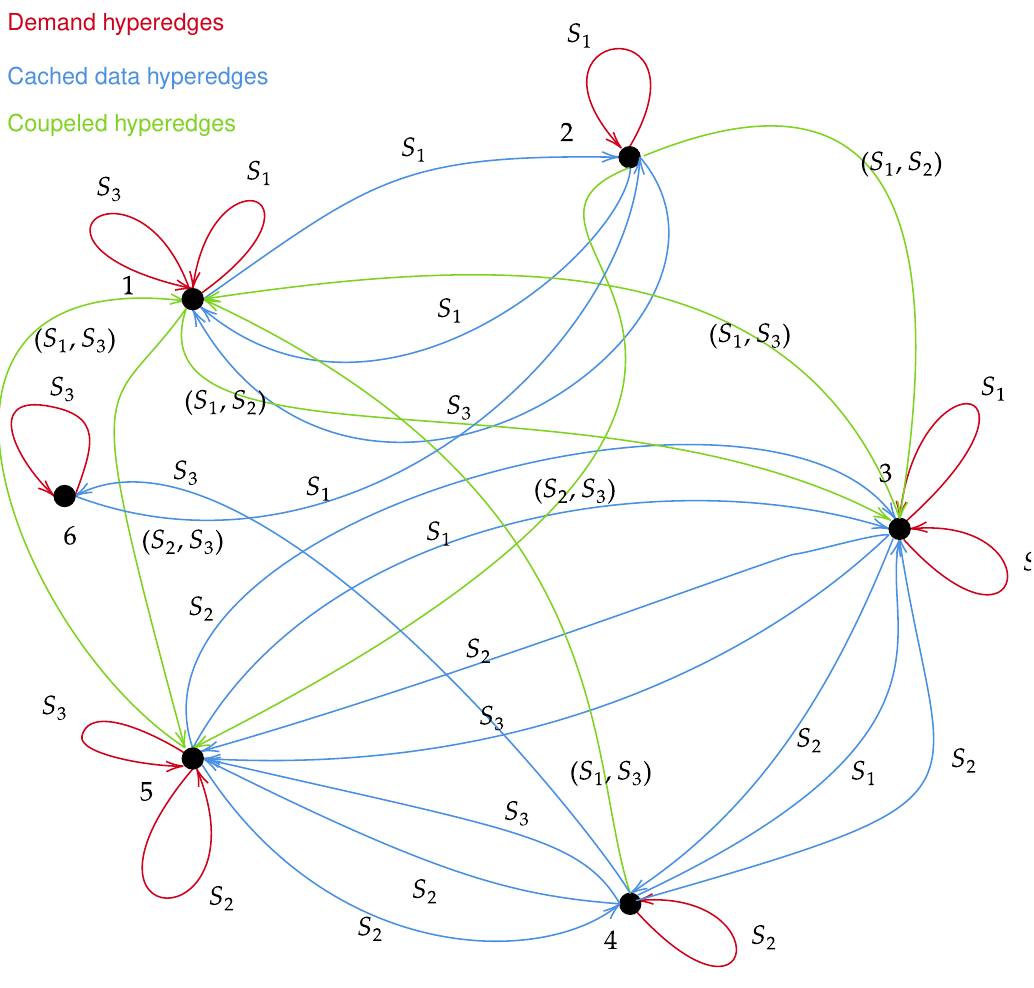}
\caption{Appendix~\ref{Example} (Illustrative Example): The side-information hypergraph of the problem setting illustrated in Figure~\ref{BigExample1}. }
    \label{BigExample2}
    \end{figure}
\begin{figure}
    \centering
\includegraphics[width=0.5\linewidth]{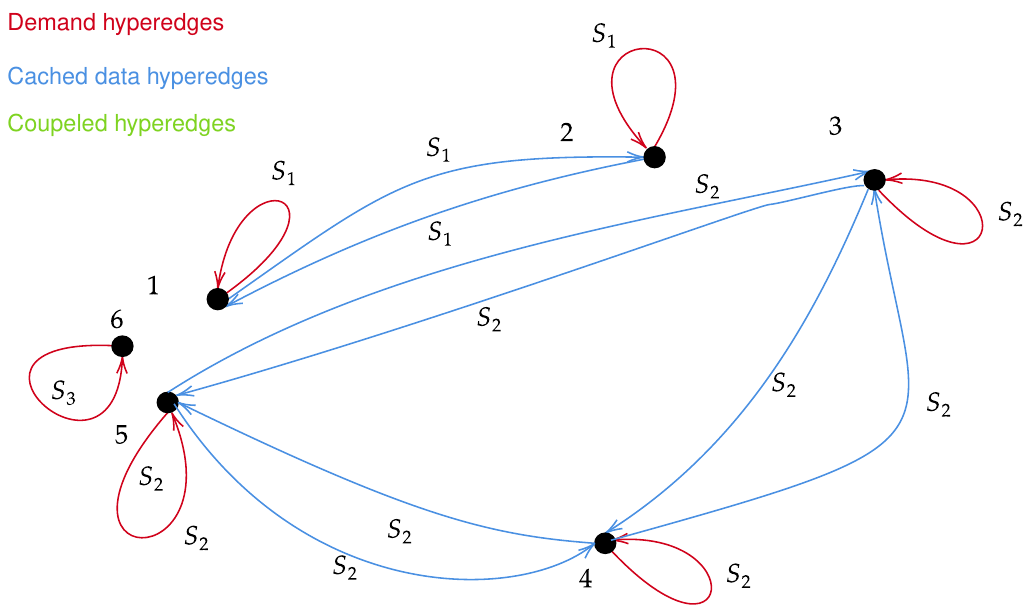}
\caption{Appendix~\ref{Example} (Illustrative Example): A valid sub-hypergraph of Figure~\ref{BigExample1} }
    \label{valid-big-subhypergraph}
    \end{figure}

We consider in particular a multi-sender index coding problem with $N=3$ senders, $K=5$ receivers, where $\mathcal{M}_1 = \{1,2,3\}, \mathcal{M}_2 = \{3,4,5\}, \mathcal{M}_3 = \{1,5,6\}$ and $\mathcal{R}(1) = \{2,3\}, \mathcal{R}(2) = \{1,3\},\mathcal{R}(3) = \{1,5\},\mathcal{R}(4) = \{2,5,6\},\mathcal{R}(5) = \{3,4\},\mathcal{R}(6) = \{2\}$ (cf. Figure~\ref{BigExample1}). Based on Definition~\ref{side-info-graph}, the corresponding side-information hypergraph is designed as shown in Figure~\ref{BigExample2}, 
   while a valid chosen sub-hypergraph of $\mathcal{G}$, is shown in Figure~\ref{valid-big-subhypergraph}. We represent it by $\mathcal{G}'$.
    The composite  adjacency matrix of $\mathcal{G}'$ is as follows:
\begin{align}\label{A'}
\mathbf{A}'=
\begin{array}{r|cccccc|cccccc|cccccc}
  & x_{1} & x_{2} & x_{3} &  x_{4} & x_{5} & x_{6} & x_{1} &x_{2} & x_{3} & x_{4} & x_{5} & x_{6}  & x_{1} & x_{2} & x_{3} & x_{4} & x_{5} & x_{6} \\ \hline
1 & \textcolor{red}{1} & \textcolor{blue}{1} & 0 & 0 & 0  & 0 & 0 & 0 & 0& 0 & 0& 0 & 0& 0 & 0& 0  & 0 & 0\\
2 & \textcolor{blue}{1} & \textcolor{red}{1} & 0 & 0 & 0  & 0 &0  & 0 & 0 & 0 & 0& 0 & 0& 0 & 0 & 0  & 0 & 0\\
3 & 0 & 0 & 0 & 0 & 0 & 0 & 0  & 0 & \textcolor{red}{1} & \textcolor{blue}{1} & \textcolor{blue}{1}& 0& 0 & 0& 0 &0 & 0  & 0 \\
4& 0 & 0 & 0 & 0 & 0 & 0 & 0& 0 & \textcolor{blue}{1} & \textcolor{red}{1} & \textcolor{blue}{1}& 0& 0 & 0& 0& 0& 0 &0\\
5 & 0 & 0 & 0 & 0 & 0 & 0 & 0& 0  & \textcolor{blue}{1} & \textcolor{blue}{1} & \textcolor{red}{1}& 0& 0 & 0& 0& 0& 0 &0\\
6 & 0 &0  & 0 & 0 & 0 & 0 & 0& 0  & 0& 0 & 0 & 0& 0 & 0& 0& 0& 0 &\textcolor{red}{1}\\
\end{array}
\end{align}
and it gives us an index code where sender one sends $x_1+x_2$, sender two sends $x_3 + x_4+x_5$, and sender three sends  $x_6$ (cf. Theorem~\ref{main-Theorem}). 

Now, to see how Theorem~\ref{Theorem-upper} works on this example, consider these three hypergraphic cliques (cf. Definition~\ref{Hypergraphic-clique})
\begin{align*}
    \mathcal{C}_1 &=\{(1,2,1,1), (1,1,1,1),(2,2,1,1), (2,1,1,1)\},\\
    \mathcal{C}_2 &=\{(3,5,2,2),(3,4,2,2), (5,3,2,2),(5,4,2,2),(4,3,2,2), (4,5,2,2)\},\\
    \mathcal{C}_3 &=\{(6,6,3,3)\}
\end{align*}
where we can indeed readily verify that $[K] = \{k,k'| (k,k',n,n') \in \cup_{j \in [3]} {\mathcal{C}_{j}}\}$. We can also readily see that $\mathcal{G}'' = \cup_{j \in [3]} {\mathcal{C}_{j}}$ is a valid  sub-hypergraph of $\mathcal{G}$ as depicted by Figure~\ref{valid-big-subhypergraph}, and   we can see that its adjacency matrix  is again $\mathbf{A}'$ in \eqref{A'}.  

  Let us now illustrate how Theorem~\ref{Theorem-lower} can be applied to the present example in order to derive a lower bound on $\hyperminrank(\mathcal{G})$. 
According to Definition~\ref{Complement}, we first construct the \emph{complementary side-information hypergraph} $\bar{\mathcal{G}}$, whose hyperedges capture all non-existing dependencies in $\mathcal{G}$. 
The adjacency matrix of this complement, denoted by $\bar{\mathbf{A}}$, encodes the structure of $\bar{\mathcal{G}}$ and is obtained by inverting the connectivity pattern of $\mathcal{G}$ while preserving its block organization across senders. 
Then, the complementary composite adjacency matrix takes the form:
\begin{align}
\mathbf{A}'=
\begin{array}{r|cccccc|cccccc|cccccc}
  & x_{1} & x_{2} & x_{3} &  x_{4} & x_{5} & x_{6} & x_{1} &x_{2} & x_{3} & x_{4} & x_{5} & x_{6}  & x_{1} & x_{2} & x_{3} & x_{4} & x_{5} & x_{6} \\ \hline
1 & \textcolor{red}{1} & 0 & 0 & 1 & 1  &  1 & 0  & 1 & 0 & 1 & 0& 1 & \textcolor{red}{1}& 1 & 1& 1  & 0 & 1\\
2 & 0 & \textcolor{red}{1} & 0 & 1 & 1  & 1 &1  & 0 & 0 & 1 & 0& 1 & 0& 0 & 1 & 1  & 0 & 1\\
3 & 0 & 1 & \textcolor{red}{1} & 1 & 1 & 1 & 1  & 1 & \textcolor{red}{1} & 0 &0& 1& 0 & 1& 0 &1& 0  & 1 \\
4& 0 & 1 & 0 & 0 & 1 & 1 & 1& 1 & 0 & \textcolor{red}{1} & 0 & 1& 0 & 1& 1& 0& 0 &0\\
5 & 0 & 1 & 0 & 1 & 0 & 1 & 1& 1  & 0 & 0  & \textcolor{red}{1}& 1& 0 & 1& 1& 1& \textcolor{red}{1} &1\\
6 & 1 &0  & 1 & 1 & 1 & 1 & 1& 1  & 1& 1 & 1 & 1& 1 & 1& 1& 1& 1 &\textcolor{red}{1}\\
\end{array}
\end{align}
    Note that the biggest directed graph clique that satisfies the hypotheses of Theorem~\ref{Theorem-lower} is a clique of the form $\mathcal{F} \triangleq\{\{(5,6),(5,5),(6,5),(6,6)\}|k \in  [5]\}$ corresponding to sender $3$.  Therefore, since $|\mathcal{V}{(\mathcal{C})}| =2, \forall \mathcal{C} \in \mathcal{F}$, directly from Theorem~\ref{Theorem-lower}, we have the converse $\hyperminrank(\mathcal{G}) \geq 2$.

\bibliographystyle{IEEEtran}
\bibliography{ref}
\end{document}